\documentclass{article}
\usepackage{iclr2027_conference,times}
\usepackage{microtype}
\usepackage{booktabs}
\usepackage{array}
\usepackage{amsmath,amssymb}
\usepackage{graphicx}
\usepackage{tikz}
\usetikzlibrary{arrows.meta,positioning,calc,fit}
\usepackage[hypertexnames=false,colorlinks=true,linkcolor=black,citecolor=teal,urlcolor=blue]{hyperref}
\usepackage{url}
\usepackage{xcolor}
\usepackage{float}
\usepackage{placeins}
\usepackage{afterpage}

\newcommand{\ourmethod}{SepGen}
\newcommand{\method}{\ourmethod}
\newcommand{\system}{\ourmethod}
\newcommand{\ckptsep}{separation checkpoint}
\newcommand{\ckptgen}{generation checkpoint}

\newsavebox{\fitbox}
\newcommand{\fitwidth}[1]{\sbox{\fitbox}{#1}%
  \ifdim\wd\fitbox>\textwidth\resizebox{\textwidth}{!}{\usebox{\fitbox}}\else\usebox{\fitbox}\fi}

\newcommand{\runinsub}[1]{\par\smallskip\noindent\refstepcounter{subsection}\textbf{\thesubsection\hspace{0.5em}#1.}\hspace{0.5em}}
\title{\ourmethod{}: Multi-Stem Audio-Video\\
\mbox{Separation} and \mbox{Generation} in a Single Model}
\author{Aviad Dahan$^{1}$ \quad Rajaei Khatib$^{1}$ \quad Yonatan Bitton$^{2}$ \quad Idan Szpektor$^{2}$ \quad Lior Wolf$^{1}$ \quad Raja Giryes$^{1}$\\
$^{1}$Tel Aviv University \quad $^{2}$Google}

\iclrfinalcopy

\begin{document}
\maketitle
\lhead{}\renewcommand{\headrulewidth}{0pt}

\begin{abstract}
A 4D audio-visual scene comprises a video, the dynamic geometry it depicts,
and the sound sources that populate it, each with its own position and
trajectory. Rendering such a scene from a novel viewpoint requires that every
source be available as an individual waveform, so that it can be localized in
the scene and propagated to the observer before the signals are mixed. Joint
audio-video generators can synthesize the video and its soundtrack, but the
soundtrack is emitted as a single audio-mix in which the sources are not
individually accessible. We present \method{}, which extends a pretrained audio-video generator to
emit the video, the mixed soundtrack, and one waveform per captioned source
in one joint sampling run.
\method{} supports two complementary modes: generation and separation. In generation mode, each source caption specifies what its stem contains. A two-speaker dialogue, for example, comes out as one stem per speaker in the original turn order. In separation mode, the input audio-mix remains clean while the captions specify what to extract, so the model can decompose a recording from a free-text description. We evaluate generation on scenes synthesized from text, and separation on scenes rendered by other generators and on real recordings of speech, music, and sound effects. Given an audio-mix and captions that carry the spoken lines, \method{} outperforms language-conditioned separators, most clearly on speech, and it keeps the lead when the lines are removed from the captions.
Code, checkpoints, and datasets are available at \url{https://sepgen.github.io/}.
\end{abstract}

\section{Introduction}
\label{sec:intro}

\begin{figure*}[t]
  \centering
  \vspace{-0.1in}
  \includegraphics[width=\linewidth]{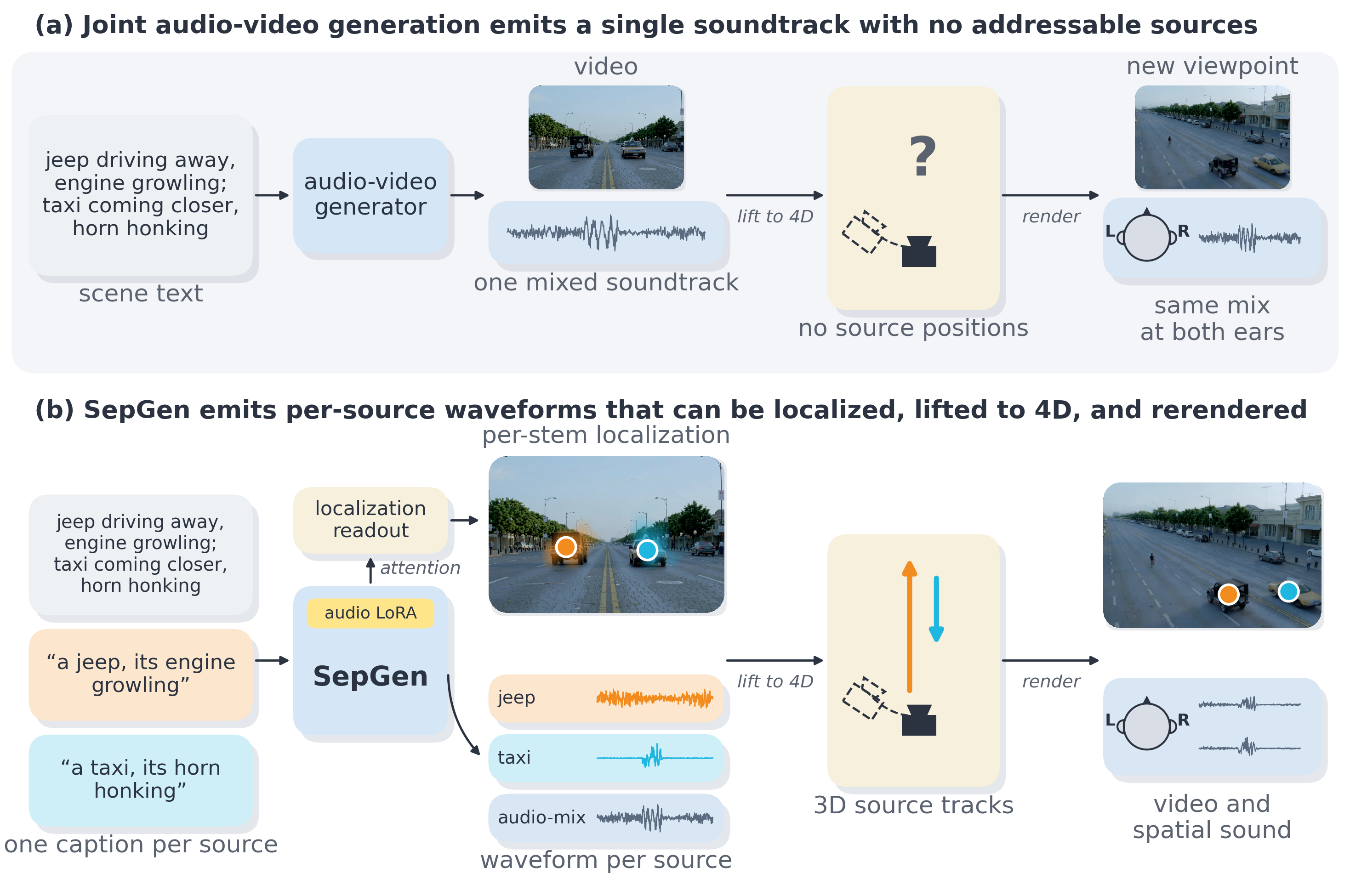}
  \vspace{-0.3in}
 \caption{\textbf{\method{} overview and motivation.}
    (a)~Joint audio-video generation emits one soundtrack with no addressable
    source: the video is lifted to 4D and rerendered from a new viewpoint, but
    both ears hear the same mix.
    (b)~\method{} emits one waveform per captioned source, each placed at its
    source in the lifted scene and rendered as spatial sound from the new
    viewpoint (illustration).
    A demonstration video is available on the project page.}
  \label{fig:teaser}
  \vspace{-0.1in}
\end{figure*}

Recent joint audio-video generators \citep{hacohen2026ltx2,ovi2025} synthesize
video with synchronized sound, but output the entire scene as one waveform,
the \emph{audio-mix}, whose sources cannot be moved, removed, or rerendered
individually. Language-queried separation
\citep{liu2024audiosep,flowsep2025,samaudio2026} can estimate a described
source afterward, but requires an additional model run and handles each
caption in isolation, which fails when two sources belong to the same
category: asked separately for each of two men speaking, it might return the
same speaker.

Per-source waveforms matter most when the scene itself is rerendered.
Monocular video methods lift a video to a 4D scene that can be viewed from a
new camera pose \citep{karhade2025any4d,litman2026lift4d,zhang2025monst3r},
and spatial audio renderers propagate each source to the observer
\citep{chen2020soundspaces,xie2025sonic4d,jin2026sonoworld,liang2023avnerf}.
An audio-mix recorded at the original pose, however, exposes no source to
place, so the sound stays fixed (Figure~\ref{fig:teaser}a).

We present \method{}, an audio-video generator that also outputs each
source: from a scene caption and one caption per source, it produces the
video, the audio-mix, and one waveform per source, a \emph{stem}, in a
single sampling run (Figure~\ref{fig:interface}). Predicting both stems
jointly from the shared audio-mix lets each follow its own caption while
sharing the scene's timing, and yields the per-source waveforms that a
spatial renderer requires (Figure~\ref{fig:teaser}b).

Adding stems to a pretrained backbone must not disturb what the backbone
already produces. \method{} therefore appends the two stems to the
backbone's audio sequence as additional channels, trained through a
low-rank audio adapter, and keeps the audio-mix as a protected channel
that attends only to itself and bypasses the adapter, so the backbone's
soundtrack and video are unchanged while the stems attend to the
audio-mix and to each other at aligned rotary positions. The stems must
also stay apart from each other: block-diagonal caption routing gives
each stem only its own caption, and Cross-Stem Attention Guidance steers
it away from its sibling's. The same model can either generate a
scene together with its stems, or separate the stems from an observed
scene. Since the backbone is a denoising model, the noise level of the
audio-mix selects between the two: separation holds the observed
audio-mix at zero noise, whereas generation denoises it together with the
stems, which then attend to its current denoised estimate (Estimated
Separation). We therefore train on separation first and briefly continue
on noisy audio-mixes, yielding one checkpoint that does both.

Our contributions are: (i) \method{}, to our knowledge the first framework
that turns a pretrained joint audio-video generator into a multi-stem
generator, emitting the video, its soundtrack, and one waveform per captioned
source in a single sampling run. (ii) Mechanisms that keep the stems
caption-faithful and the pretrained output intact: a protected audio-mix
channel that bypasses the adapter, aligned rotary positions, block-diagonal
caption routing, and a schedule in which the stems attend to the estimated clean
audio-mix. (iii) A single set of weights for both modes, selected by the
audio-mix noise level: a separation-trained checkpoint already generates
stems, and a brief continuation reduces leakage between the generated
stems, improves speech intelligibility, and stabilizes results across
seeds, while preserving separation. We evaluate both modes on generated, cross-generator, and real
scenes, with automatic metrics and a human A/B study.

\afterpage{
\begin{figure*}[t]
  \centering
\vspace{-0.1in}
  \includegraphics[width=\linewidth]{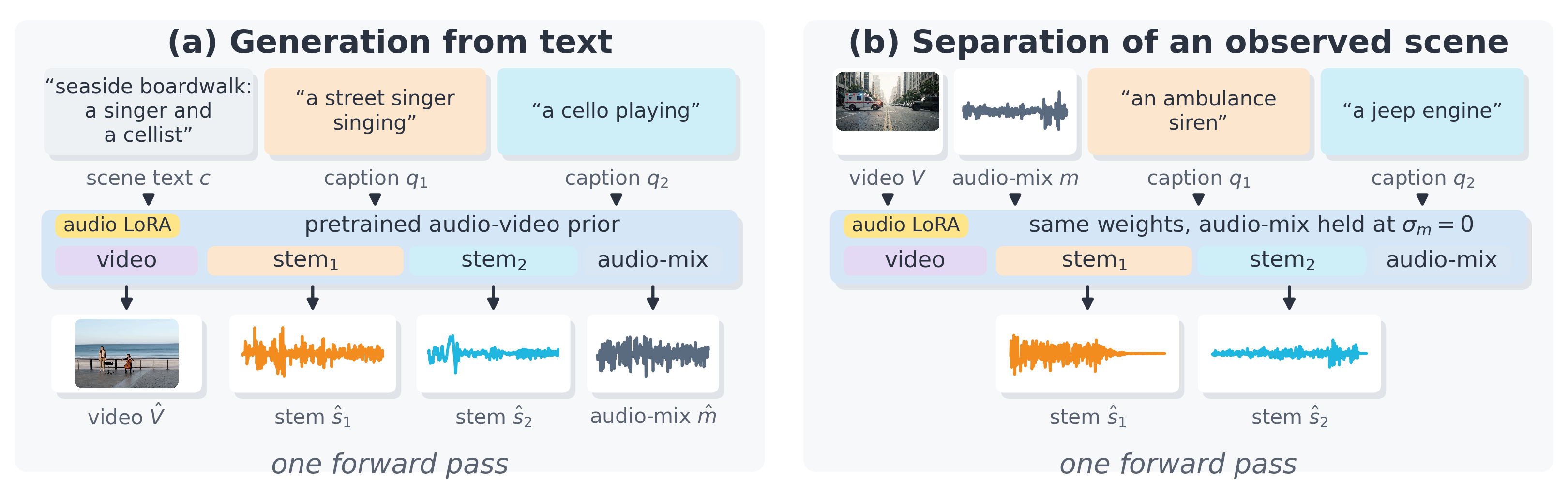}
  \vspace{-0.2in}
  \caption{\textbf{\method{}: a unified representation for
    multi-stem generation and separation.}
    (a)~Generation: scene text and two source captions yield the video,
    audio-mix, and per-source waveforms.
    (b)~Separation: from an observed video and audio-mix, the same model
    extracts both captioned stems.}
  \label{fig:interface}
  \vspace{-0.1in}
\end{figure*}
}

\section{Related Work}
\label{sec:related}

\noindent\textbf{Joint audio-video generation.} Joint audio-video generators synthesize
a video and its synchronized soundtrack from text, either by coupling two
modality streams, as in LTX-2 \citep{hacohen2026ltx2}, our backbone's family,
and Ovi \citep{ovi2025}, or by bridging separately pretrained generators
\citep{seeingandhearing2024}. Proprietary systems such as Veo~3.1 Lite
\citep{veo31lite2026}, which renders our Veo benchmark, also generate video
and audio jointly. Video-to-audio (V2A) models add a soundtrack to an
existing video
\citep{moviegen2024,mmaudio2024,klingfoley2025,hunyuanvideofoley2025,vaura2024,foleyomni2026}.
All of them output one audio-mix per scene, whereas \method{} outputs a
separate waveform for each source, in both of its modes. Like prior LoRA
adaptations of such generators \citep{hu2022lora,iclora2024,idlora2026},
\method{} adapts its backbone, but its extra audio tracks are captioned stems
that neither the video nor the audio-mix attends to, and the audio-mix
bypasses the LoRA (Section~\ref{sec:graph}). Its Cross-Stem Attention
Guidance (Section~\ref{sec:guidance}) adapts Normalized Attention Guidance
\citep{nag2025}, with the sibling caption as the negative prompt.

\noindent\textbf{Multi-source audio generation.} Multi-source models emit several stems at once. In
music, MSDM \citep{msdm2023} and its latent and text-conditioned
extensions \citep{karchkhadze2025msgld,gmsdi2024} generate and separate
stems, and other models generate, add, or text-specify stems
\citep{msldm2024,jen1composer2023,musicgenstem2025,stemgen2023,stemphonic2026}. Closest to ours, MGE-LDM \citep{mgeldm2025} switches between
generation and extraction by choosing which tracks to denoise, like our
two modes, but is audio-only and extracts one source per run. Given a
video, other models generate one clip per detected source, speech and
background tracks, one track per event, or only a text-selected source
\citep{dagli2024see2sound,dualdub2025,stepbystepv2a2025,selva2025}, or
fuse detected sources into one audio-mix \citep{ss2a2024}. To our
knowledge, no method generates a video with several captioned stems in
one denoising run.

\noindent\textbf{Language-queried source separation.} Query-conditioned separation
extracts from an audio-mix only the source specified by a query, such as
text, a visual region, or an audio example. Language-queried audio source
separation (LASS) uses free-form text \citep{liu2022lassnet,dong2023clipsep}.
Our \emph{cascade} baselines (Section~\ref{sec:setup}) are three released LASS
models. AudioSep \citep{liu2024audiosep} conditions a separation network on
CLAP text embeddings \citep{wu2023laionclap}, FlowSep \citep{flowsep2025} is a
rectified-flow generative separator, and SAM~Audio \citep{samaudio2026} is a
flow-matching transformer that also accepts visual-mask and time-span prompts.
We do not compare with the recent FlowSep~2 \citep{yuan2026flowsep2}, whose
models were unreleased at the time of writing. Other separators vary the query
encoder or network \citep{ma2024clapsep,wang2025soloaudio}, parse sources with
a large language model \citep{mahmud2024opensep}, combine text, image, and
audio queries \citep{cheng2024omnisep}, extract several sources in one pass
but only from fixed categories \citep{saijo2024tuss,paissan2025fastuss}, or
reuse pretrained text-to-audio diffusion models without separation training
\citep{zerosep2025,lee2025dgmo}.

\noindent\textbf{Audio-visual and generative separation.} Targets can also be specified by sounding objects or musicians' motion
\citep{zhao2018soundofpixels,gao2019coseparation,gan2020musicgesture},
speakers' faces \citep{ephrat2018looking,gao2021visualvoice}, or a typed
speaker description \citep{hao2023llmtse}. Generative separators exist
for speech \citep{scheibler2023diffsep}, audio-visual mixes
\citep{huang2023davis}, and film audio with fixed speech, music, and
effects stems \citep{avcass2026}. MMAudioSep \citep{mmaudiosep2026}
turns a video-to-audio model into a separator that can still generate,
but extracts one source per pass. Since query-based separators process
each query independently, queries for two speakers can both return the
louder voice (Section~\ref{sec:intro}); \method{} instead predicts both
stems jointly from the shared audio-mix.

\section{Method}
\label{sec:method}

\method{} extends a joint audio-video generator so that, in
addition to the video and its soundtrack, it outputs one waveform per
sound source. This enables a 4D audio-visual scene renderer to place each source in space over time (Figure~\ref{fig:teaser}). In addition to generation, \method{} can act as a source separator given an existing video. In generation, the model synthesizes the
video, the audio-mix, and the stems from a conditioning text. In separation, the video and audio-mix are inputs and \method{} extracts the captioned
stems. 
Section~\ref{sec:training} describes the two-stage training and Section~\ref{sec:schedule} the inference.

\noindent \textbf{Problem Formulation.} Let \(F\) denote a flow matching diffusion audio-video generator with two transformer streams, one for the video tokens and one for the audio tokens, e.g.
LTX-2.5 \citep{hacohen2026ltx2}.
In every block, each stream self-attends over its own tokens and cross-attends to
the conditioning scene caption \(c\). Additionally, the two streams exchange information
through bidirectional audio-video cross-attention. Denoising both
latents jointly and decoding them yields a video \(V\) and its audio-mix
\(m\), \((V,m)=F(c)\). For brevity, \(V\) denotes both the video and its
latent.

\method{} is formulated for scenes with \(N\) sound sources, described by
per-source captions \(Q=(q_1,\dots,q_N)\); we train and evaluate \(N{=}2\).
Our model \(F_\theta\) keeps both streams of \(F\) and
represents each source as additional tokens in the audio stream.
The audio VAE encoder maps each stem and the audio-mix to
latents \(z_{s_1}\), \(z_{s_2}\), and \(z_m\), each with the temporal
length of the pretrained audio latent, and its decoder maps each latent
back to a waveform. The audio stream receives the three latents as one
token sequence, \(z_a=[z_{s_1}\mid z_{s_2}\mid z_m]\), concatenated along
the token axis. Each channel keeps the temporal positions of the
pretrained audio latent, so the \(j\)-th token of every channel refers to
the same time instant in the audio stream. Each stem channel attends to itself, the
other stems, the audio-mix, the video, and  only to its own
caption \(q_i\). The audio-mix channel attends to \(c\) as in
\(F\). Neither the audio-mix nor the video attend to the stems, so
information flows only from the scene to the stems. A low-rank
adapter (LoRA) \citep{hu2022lora} with parameters \(\theta\) updates the
projections of the audio stream on the stem tokens only, while the weights of \(F\) stay frozen. 
Each forward pass of \(F_\theta\) predicts the velocities of the video and of all three audio
channels, and the sampling schedule of each mode determines which of
these predictions are used.
Section~\ref{sec:graph} details this layout.
\method, \(F_\theta\), is a single model that supports two modes:
\begin{equation}
  \text{Generation: } (\hat V,\hat m,\hat s_1,\hat s_2)=F_\theta(c,Q),
  \qquad
  \text{Separation: } (\hat s_1,\hat s_2)=F_\theta(c,Q,V,m).
  \label{eq:modes}
\end{equation}
In generation (Figure~\ref{fig:interface}a), \(F_\theta\) samples the
video, the audio-mix, and one stem per source from the captions alone.
In separation (Figure~\ref{fig:interface}b), \(F_\theta\) receives an
observed video \(V\) and audio-mix \(m\) and returns only the stems. The
two modes differ in how \(F_\theta\) treats \(z_V\) and \(z_m\): separation keeps
them clean through sampling, and generation denoises
them together with \([z_{s_1}\mid z_{s_2}]\).

\runinsub{Two-Stage Training}\label{sec:training}

\textbf{Training data.}
Each training example consists of a video, its audio-mix, the waveform
of each of its two sources, a scene caption, and a caption per source.
We assemble 4{,}843 such examples from four origins: CelebV-HQ
\citep{zhu2022celebvhq}, URMP \citep{li2018urmp}, two-source scenes
that we generate, and synthetic audio-mixes that combine clips from
different VGGSound \citep{chen2020vggsound} videos and MUSIC-21
\citep{zhao2019soundofmotions} solo performances. The two sources of an example may sound at the same time
or alternate, as in a dialogue where the speakers take turns. Each
example has up to three sets of captions at different verbosity levels, and each training step draws
one set at random. Appendix~\ref{sec:traindata} describes this process in detail.

\textbf{Objective.}
We train LoRA parameters \(\theta\), with the
flow-matching objective of \(F\). For a clean latent \(z\) and a noise
level \(\sigma\in[0,1]\), the noisy latent and its target velocity are
\begin{equation}
  z^{\sigma}=(1-\sigma)\,z+\sigma\,\varepsilon,
  \qquad
  u=\varepsilon-z,
  \qquad
  \varepsilon\sim\mathcal N(0,I).
  \label{eq:flow}
\end{equation}
The two stem latents are noised to a common level \(\sigma_s\)
and the audio-mix latent is noised to a level $\sigma_m$ $\le\sigma_s$. The
distribution of \(\sigma_m\) is the only difference between the two
stages.
Thus, the input to \(F_\theta\) is
\begin{equation}
  \mathcal I(\sigma_s,\sigma_m)=\big([z_{s_1}^{\sigma_s}\mid z_{s_2}^{\sigma_s}\mid
  z_m^{\sigma_m}],\,V,\,c,\,Q\big).
  \label{eq:input}
\end{equation}
It provides \(F_\theta\) with the source timestep \(\sigma_s\) and the
audio-mix timestep \(\sigma_m\). Let \(v_{\theta,i}\) denote the velocity
that \(F_\theta\) predicts on the tokens of stem \(s_i\), and \(u_{s_i}\)
the target velocity of Eq.~\eqref{eq:flow} for \(z_{s_i}\). The loss is
\begin{equation}
  \mathcal L=\sum_{i=1}^{2}
  \mathbb E\Big[\big\lVert v_{\theta,i}\big(\mathcal I(\sigma_s,\sigma_m)\big)
  -u_{s_i}\big\rVert_2^2\Big],
  \label{eq:loss}
\end{equation}
where the expectation is over training examples. Only the stem tokens receive the loss. The predictions on the
video and audio-mix tokens are not supervised, and these tokens serve as
inputs that tell the stems what the scene contains.

\textbf{Stage 1: separation.}
The first stage trains \(F_\theta\) for the separation mode of
Eq.~\eqref{eq:modes}: given an observed video, its audio-mix, and
the two source captions, predict the two stems. The LoRA update starts
at zero, so training begins from the behavior of \(F\). We set
\(\sigma_m{=}0\) for every example, so the audio-mix is always clean and
the stems learn to decompose it into the sources described by \(q_1\)
and \(q_2\). This stage trains only the separation task, yet the resulting checkpoint
can already generate stems by attending to the estimated clean audio-mix at
every step (Appendix Table~\ref{tab:schedule}); Stage 2 improves
this ability.

\textbf{Stage 2: generation.}
In the generation mode of Eq.~\eqref{eq:modes}, no clean audio-mix
is available: the audio-mix is denoised together with the stems, so the
stems must attend to an audio-mix that is still noisy. In both modes, the
stems solve the same task, decomposing the audio-mix they attend to into the
captioned sources. The modes differ only in whether that audio-mix is
clean or still being denoised. The second stage therefore initializes
\(\theta\) from the \ckptsep{} and continues training with the same data
and loss, changing only how \(\sigma_m\) is drawn. Concretely, we first
sample an auxiliary audio-mix noise level \(\sigma'_m\) (independently of
\(\sigma_s\)) and then define \(\sigma_m\) as
\begin{equation}
  \sigma_s,\sigma'_m \overset{\mathrm{ind}}{\sim}p(\sigma),
  \qquad b\sim\operatorname{Bernoulli}(0.3),
  \qquad \sigma_m=(1-b)\min(\sigma'_m,\sigma_s).
  \label{eq:continuation}
\end{equation}
With probability 0.3 (\(b{=}1\)), the audio-mix stays clean as in
Stage~1, which preserves the separation mode. Otherwise (\(b{=}0\)), the
audio-mix is noised to level \(\sigma_m=\min(\sigma'_m,\sigma_s)\), i.e.,
capped at
\(\sigma_s\). Since \(\sigma'_m\ge\sigma_s\) in half of the draws, half
of these samples place the audio-mix at exactly the stem level. The cap
keeps the audio-mix no noisier than the stems, which covers the two
conditions that occur during generation: an audio-mix at the same noise
level as the stems, and a clean estimate of the audio-mix
(Section~\ref{sec:schedule}). This stage yields the \ckptgen{}, which
performs both separation and generation. Compared with the separation
checkpoint used as a generator, the continuation raises J-swap from 1.95
to 2.07, lowers own-line word error from 0.10 to 0.07, and reduces the
seed spread of J-swap from 0.13 to 0.01 (Appendix
Table~\ref{tab:schedule}).

\textbf{Stem Layout and Routing.}\label{sec:graph}
In order not to degrade the base model's capabilities, we let each stem attend to the audio-mix, but block any information flow back into the mix.
We concatenate the audio tokens as
\(z_a=[z_{s_1}\mid z_{s_2}\mid z_m]\), keeping each channel's pretrained temporal shape.
We enforce this routing with the attention mask
\begin{equation}
  \Lambda=\begin{bmatrix}
    1 & 1 & 1 \\
    1 & 1 & 1 \\
    0 & 0 & 1
  \end{bmatrix},
  \label{eq:allow}
\end{equation}
(rows are queries, columns are keys, ordered as \((s_1,s_2,m)\)).
Thus stems can attend to stems and the mix, while the mix attends only to itself.

Attention alone is not enough: LoRA would still change mix-token projections.
We therefore gate the audio-side LoRA update off for mix tokens (indices \(\mathcal J_m\)):
\begin{equation}
  h'_j=W_0h_j+g_jBA(h_j),\qquad
  g_j=\mathbf 1[j\notin\mathcal J_m].
  \label{eq:lora-gate}
\end{equation}
Mix tokens use the frozen projection \(W_0\) (\(g_j{=}0\)), stem tokens use the adapted one (\(g_j{=}1\)). For caption routing, block-diagonal cross-attention lets \(s_i\) attend only to \(q_i\) and \(z_m\) only to \(c\).
We use the same rotary positions so the \(j\)-th token of every channel aligns in time.
Overall, neither attention nor adapted weights can carry information from stems into the mix, only stem tokens are updated.%

\runinsub{Inference Schedules}\label{sec:schedule}

\begin{figure*}[t]
  \centering
  \includegraphics[width=\linewidth]{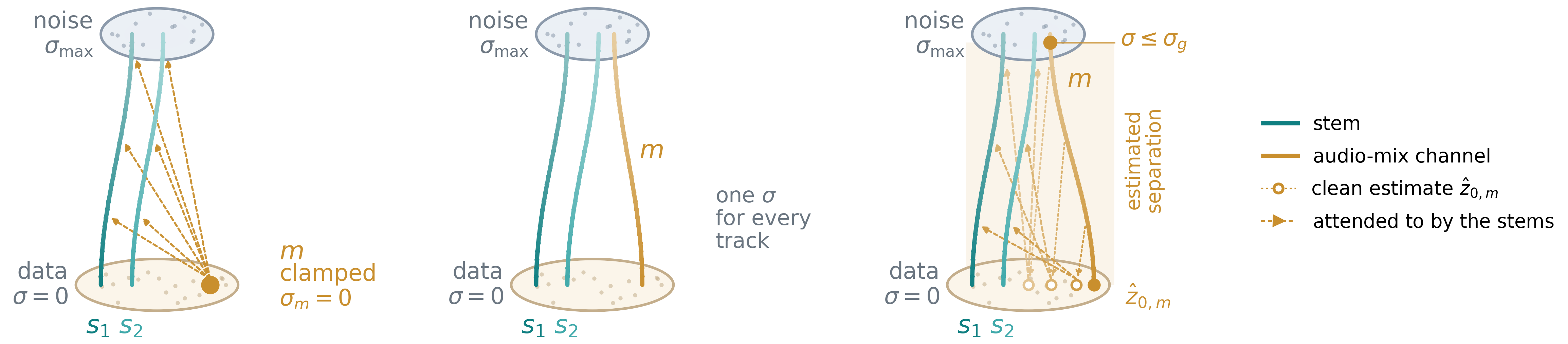}
\vspace{-0.2in}

  \caption{\textbf{Inference schedules in \(\sigma\)-space.} Left to right:
    (a)~Separation keeps the audio-mix \(m\) clean. (b)~Shared-noise generation, the naive approach, denoises the stems and the audio-mix at one shared noise level. (c)~\method{}'s stems attend to noisy \(m\) above
    \(\sigma_g\), then its clean estimate \(\hat z_{0,m}\) below it.
    }
  \label{fig:schedule}
  \vspace{-0.1in}
\end{figure*}

At each sampling step \(k\), with the noise level \(\sigma_k\) decreasing
from 1 to 0, \(F_\theta\) predicts a velocity for every latent. The audio stream predicts the
stem velocities \(\hat v_{s_i}\) and the audio-mix velocity \(\hat v_m\),
and the video stream predicts the video velocity \(\hat v_V\). A latent being denoised is
updated by a solver step:
\begin{equation}
  z^{\sigma_{k+1}}=z^{\sigma_k}+(\sigma_{k+1}-\sigma_k)\,\hat v .
  \label{eq:euler}
\end{equation}
The two modes differ in which latents are updated
(Figure~\ref{fig:schedule}).

\textbf{Separation.}
Only the stems are denoised, starting from noise. At every step,
\(F_\theta\) receives the input \(\mathcal I(\sigma_s,0)\) of
Eq.~\eqref{eq:input} with \(\sigma_s{=}\sigma_k\): the audio-mix is the
observed clean \(z_m\). The stems are updated with
Eq.~\eqref{eq:euler} as illustrated in Figure~\ref{fig:schedule} (a).

\textbf{Generation.}
During generation, all latents are being denoised together, video and audio-mix are updated
with their own predicted velocities. If the stems attend to the noisy
audio-mix throughout sampling, they may produce  content that
the final audio-mix does not contain. Instead, they can attend to the estimated
audio-mix that the current step predicts, which follows from
Eq.~\eqref{eq:flow} as \(\hat z_{0,m}=z_m^{\sigma_k}-\sigma_k\hat v_m\).
This estimate is unreliable at high noise levels, so we switch between
the two at a threshold \(\sigma_g\):
\begin{equation}
  z_m^{\mathrm{read}}(\sigma_k)=
  \begin{cases}
    z_m^{\sigma_k}, & \sigma_k>\sigma_g,\\
    \hat z_{0,m}, & \sigma_k\le\sigma_g,
  \end{cases}
  \label{eq:inference-schedule}
\end{equation}
with \(\sigma_g{=}0.97\), so in the first denoising steps the stems attend to the
noisy audio-mix.
We call this scheme \emph{Estimated Separation} (Figure~\ref{fig:schedule}c).
Stage~2
training covers both regimes: its samples with \(\sigma_m{=}\sigma_s\)
match the noisy steps, and its clean samples match Estimated
Separation. Appendix Table~\ref{tab:schedule} compares this schedule
with alternatives.

\textbf{Guidance.}\label{sec:guidance}
In both operation modes, the stems use our introduced Cross-Stem Attention
Guidance (CSAG), which applies Normalized Attention Guidance
\citep{nag2025} inside each stem's caption cross-attention with the
sibling caption as the negative branch, at scale \(\lambda{=}2\)
(Appendix~\ref{sec:csag}).

\section{Experiments}
\label{sec:experiments}

We evaluate both modes of \method{} against cascade baselines that apply
a language-queried separator to the same audio-mix. We address two questions: does the
generation mode produce stems that adhere to their captions without
cross-stem leakage (Section~\ref{sec:generation}), and does the same
model separate real and externally generated scenes while retaining this
ability after generation training (Section~\ref{sec:separation})? We
then report a user study (Section~\ref{sec:userstudy}) and ablations
(Section~\ref{sec:ablations}).

\subsection{Setup}
\label{sec:setup}

\noindent \textbf{Benchmarks and baselines.} \textbf{Generation.} For the generation task, we curate 194
two-source scene prompts (134 sound scenes and 60 two-speaker
dialogues) using Claude for a draft and manually approving each prompt. For each prompt, the \ckptgen{} generates the video, the
audio-mix, and one stem per source caption; each baseline then receives
this generated audio-mix and separates the same two captions from it.
\textbf{Separation.} For separation, we curate two benchmarks of rendered
scenes: 149 scenes rendered by Veo~3.1 Lite 
and 64 scenes rendered by LTX-2.3. We further evaluate on 172 real
recordings from SAM-Audio-bench, each with a single queried source. In this benchmark, \method{} receives the queried source as \(q_1\) and a fixed caption
describing the residual background as \(q_2\), the swap margins score
the target stem against the list of the other sound events annotated in
the clip, and the speech recordings, which have no transcripts, are
scored with Judge and CLAP. In
every case, the input is an existing video with its audio-mix, and
every method, including both checkpoints of \method{}, extracts the
captioned sources from it. We compare with FlowSep \citep{flowsep2025}, AudioSep
\citep{liu2024audiosep}, and the two released SAM~Audio checkpoints
\citep{samaudio2026}, SAM large and SAM large-tv. On the Veo benchmark, we also prompt
SAM large-tv with SAM3 \citep{carion2025sam3} video masks of each source
instead of text (SA\(\times\)SAM3).

\noindent \textbf{Metrics.} On sounds, we use the SAM~Audio Judge
\citep{samaudio2026}, which rates a stem from 1 to 5 given its
description and the input audio-mix, and the CLAP similarity
\citep{wu2023laionclap,xiao2024claplass} between a stem and its
description. Since CLAP scores a copied audio-mix close to a separated
stem (Appendix Table~\ref{tab:gen3seed}), we also report swap
margins (J-swap, C-swap), the score against the stem's description
minus that against the sibling's, on which a copy scores zero. On
speech, we use Whisper \citep{radford2023whisper} word error rate
against the speaker's quoted line (own-line WER) and the cosine between
the two stems' WavLM speaker embeddings \citep{chen2022wavlm}.
Since no benchmark provides reference stems, all metrics compare each
stem with its caption. On 129 held-out scenes that do have reference
stems, \method{} outperforms every baseline in envelope correlation and
log-spectral distance (Appendix~\ref{sec:refstem}). Definitions and controls are in Appendix~\ref{sec:metrics}. In
all tables, the best value is in bold and the second underlined.

\noindent \textbf{Implementation details.} We train LoRA adapters (327M
parameters) on the audio stream of the frozen LTX-2.5 22B backbone: the
\ckptsep{} for 12{,}000 updates and the \ckptgen{} for 3{,}000 more, on
a single NVIDIA A6000. Results are averaged over three seeds, or five on
SAM-Audio-bench (Appendices~\ref{sec:settings} and~\ref{sec:seeds}).

\subsection{Generation results}
\label{sec:generation}

\begin{table*}[t]
  \centering
  \caption{\textbf{Generation benchmark.} Comparison of \method{} with cascade baselines, each of which separates the same generated audio-mix given the same captions (sounds \(n{=}134\), speech \(n{=}60\)). Both \method{} rows use the \ckptgen{}.}
  \label{tab:gen}
  \setlength{\tabcolsep}{2.1pt}
  \resizebox{\textwidth}{!}{%
  \begin{tabular}{@{}p{0.33\textwidth}*{6}{>{\raggedleft\arraybackslash}p{0.137\textwidth}}@{}}
    \toprule
    & \multicolumn{4}{c}{Sounds (\(n{=}134\))} & \multicolumn{2}{c}{Speech (\(n{=}60\))}\\
    \cmidrule(lr){2-5}\cmidrule(lr){6-7}
        Method & Judge \(\uparrow\) & J-swap \(\uparrow\) & CLAP \(\uparrow\) & C-swap \(\uparrow\) & WER \(\downarrow\) & Cosine \(\downarrow\)\\
    \midrule
        FlowSep & \(3.35\) & \(1.26\) & \(0.14\) & \(0.10\) & \(1.03\) & \(0.97\)\\
        AudioSep & \(3.23\) & \(1.27\) & \(0.14\) & \(0.11\) & \(0.91\) & \(0.87\)\\
        SAM large-tv & \(2.96\) & \(0.41\) & \(0.05\) & \(0.03\) & \(0.87\) & \(0.95\)\\
        SAM large & \(2.93\) & \(0.42\) & \(0.04\) & \(0.03\) & \(0.81\) & \(0.95\)\\
    \midrule
        Ours (joint run) & \(\underline{3.82}\) & \(\underline{2.07}\) & \(\underline{0.15}\) & \(\underline{0.18}\) & \(\mathbf{0.07}\) & \(\mathbf{0.66}\)\\
        Ours (self-separation) & \(\mathbf{4.21}\) & \(\mathbf{2.29}\) & \(\mathbf{0.16}\) & \(\mathbf{0.20}\) & \(\underline{0.10}\) & \(\underline{0.67}\)\\
    \bottomrule
  \end{tabular}}
  \vspace{-0.1in}
\end{table*}

\begin{figure}[tb]
    \centering

    \includegraphics[trim=0 122 0 150, clip, width=\linewidth]{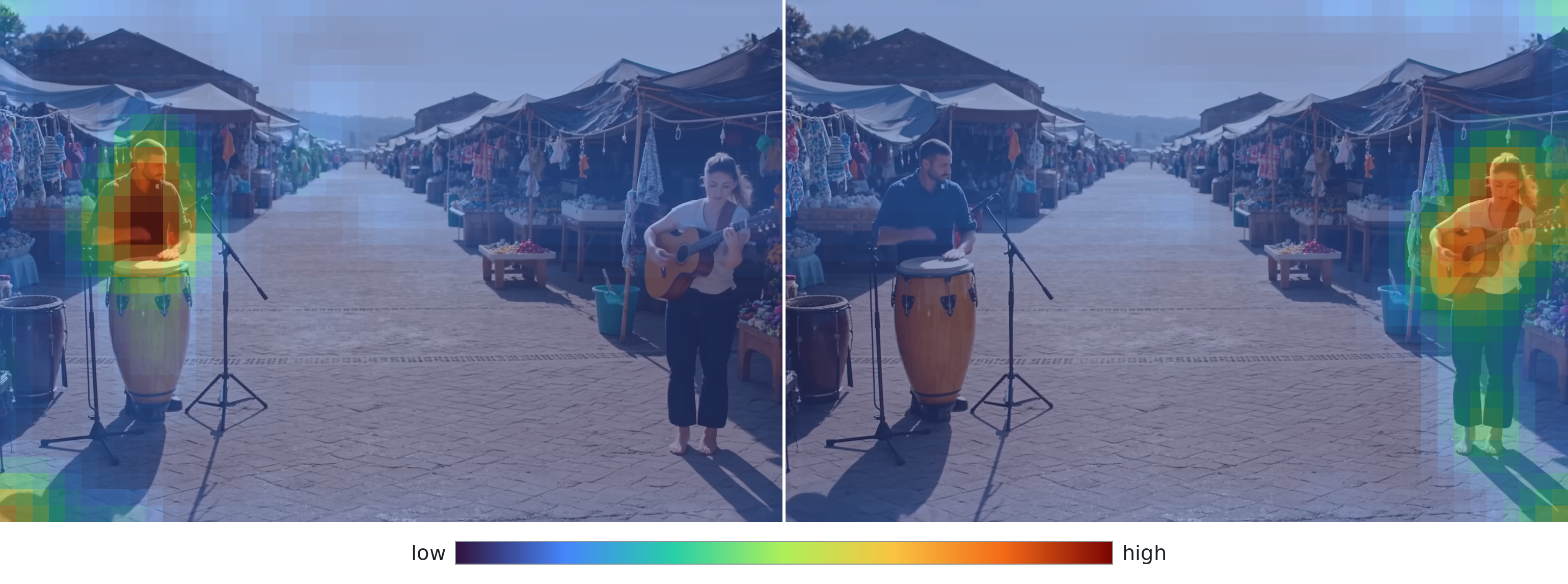}
      \vspace{-0.2in}

    \caption{\textbf{Stem-to-video attention} in a generated frame. Each stem attends
    to the source named in its caption: bongo drum (left) and acoustic guitar (right);
    warmer is higher.}
    \label{fig:stemattn}
    \vspace{-1em}
  \end{figure}

Table~\ref{tab:gen} reports \method{} in two configurations: the
stems of the generation run itself (joint run), and a second run of
the same weights in separation mode over the generated video and
audio-mix, with the same captions (self-separation).
The joint-run stems lead every cascade baseline on every metric,
significantly except for CLAP against FlowSep and against AudioSep,
whose query encoder is the scorer's CLAP checkpoint (Holm-corrected
paired tests, Appendix Table~\ref{tab:gen3seed}). The widest lead is in
J-swap: the two stems are generated jointly and each is bound to its
caption, whereas a cascade answers each query in isolation and can
return the same source twice. Self-separation raises every sounds
column at the cost of a second sampling run, while the joint-run stems
keep a lower own-line word error, within seed variation. The summed
stems also match the audio-mix more closely than those of any cascade
(Appendix Table~\ref{tab:sisdr}).
In the generated scene of Figure~\ref{fig:stemattn},
the video attention of each stem concentrates on the source named in
its caption.
Per-category results are in Appendix Table~\ref{tab:genblocks} and
qualitative examples in Appendix Figure~\ref{fig:qualgen}.

\subsection{Separation results}
\label{sec:separation}

\begin{table*}[t]
  \centering
  \caption{\textbf{Separation on the Veo benchmark} (149 scenes rendered by Veo~3.1 Lite).
  Speech cells report results with and without the quoted lines.}
  \label{tab:sep}
  \setlength{\tabcolsep}{3.3pt}
  \resizebox{\textwidth}{!}{%
  \begin{tabular}{@{}p{0.37\textwidth}*{4}{>{\raggedleft\arraybackslash}p{0.115\textwidth}}*{2}{>{\raggedleft\arraybackslash}p{0.137\textwidth}}@{}}
    \toprule
    & \multicolumn{4}{c}{Sounds (\(n{=}104\))}
    & \multicolumn{2}{c}{Speech (\(n{=}45\))}\\
    \cmidrule(lr){2-5}\cmidrule(lr){6-7}
    Method & Judge \(\uparrow\) & J-swap \(\uparrow\) & CLAP \(\uparrow\) & C-swap \(\uparrow\) & WER \(\downarrow\) & Cosine \(\downarrow\)\\
    \midrule
    AudioSep & \(3.46\) & \(1.60\) & \(\underline{0.19}\) & \(\underline{0.16}\) & \(0.93\) / \(1.03\) & \(\underline{0.89}\) / \(0.86\)\\
    FlowSep & \(3.21\) & \(1.33\) & \(0.15\) & \(0.10\) & \(0.98\) / \(0.99\) & \(0.94\) / \(0.94\)\\
    SAM large & \(3.14\) & \(0.50\) & \(0.07\) & \(0.04\) & \(\underline{0.74}\) / \(0.73\) & \(0.95\) / \(0.94\)\\
    SAM large-tv & \(3.06\) & \(0.35\) & \(0.07\) & \(0.04\) & \(0.81\) / \(0.79\) & \(0.96\) / \(0.94\)\\
    SA\(\times\)SAM3 & \(3.07\) & \(0.27\) & \(0.08\) & \(0.03\) & \(0.87\) / -- & \(0.94\) / --\\
    \midrule
    \method{} (\ckptsep{}) & \(\underline{4.36}\) & \(\mathbf{2.61}\) & \(\underline{0.19}\) & \(\mathbf{0.23}\) & \(\mathbf{0.05}\) / \(\mathbf{0.50}\) & \(\mathbf{0.61}\) / \(\mathbf{0.53}\)\\
    \method{} (\ckptgen{}) & \(\mathbf{4.39}\) & \(\underline{2.59}\) & \(\mathbf{0.20}\) & \(\mathbf{0.23}\) & \(\mathbf{0.05}\) / \(\underline{0.55}\) & \(\mathbf{0.61}\) / \(\underline{0.54}\)\\
    \bottomrule
  \end{tabular}}
  \vspace{-0.1in}
\end{table*}

\begin{table}[t]
  \centering
  \caption{\textbf{Separation on SAM-Audio-bench}. All methods are queried with identical captions.}
  \label{tab:sambench}
  \setlength{\tabcolsep}{6pt}
  \resizebox{\textwidth}{!}{%
  \begin{tabular}{@{}p{0.35\textwidth}*{4}{>{\raggedleft\arraybackslash}p{0.183\textwidth}}@{}}
    \toprule
        Method & Judge \(\uparrow\) & J-swap \(\uparrow\) & CLAP \(\uparrow\) & C-swap \(\uparrow\)\\
    \midrule
        SAM large-tv & \underline{3.12} & 0.78 & 0.13 & 0.03\\
        AudioSep & 2.81 & 0.82 & \underline{0.20} & 0.07\\
        SAM large & 2.75 & 0.63 & 0.10 & 0.01\\
        FlowSep & 2.64 & 0.51 & 0.16 & 0.03\\
    \midrule
        \method{} (\ckptsep{}) & \textbf{3.70} & \underline{1.42} & \textbf{0.21} & \underline{0.11}\\
        \method{} (\ckptgen{}) & \textbf{3.70} & \textbf{1.52} & \textbf{0.21} & \textbf{0.13}\\
    \bottomrule
  \end{tabular}}
  \vspace{-0.1in}
\end{table}

Tables~\ref{tab:sep} and~\ref{tab:sambench} report separation on the
Veo benchmark and SAM-Audio-bench, and Appendix Table~\ref{tab:ltx}
on the LTX-2.3 benchmark, for both checkpoints, which perform
comparably on every benchmark.
\textbf{Veo benchmark.} Both checkpoints surpass every cascade baseline
in Table~\ref{tab:sep} on Judge and both swap margins, and prompting
SAM large-tv with SAM3 video masks instead of text (SA\(\times\)SAM3)
does not close the gap. A cascade's query encoder cannot exploit the
quoted line in a speech caption, so each speech cell also reports every
method with all quoted lines removed (second value); the lead persists.
Without the lines, 4 to 5 of the 90 stems fall silent, which partly
explains why speaker cosine improves while word error rises. A dedicated
speech separator, SepFormer \citep{Subakan2020AttentionIA}, matches the
\ckptsep{} only when the quoted lines are removed (Table~\ref{tab:sep} and
Appendix Table~\ref{tab:speechsep}). SAM~Audio's released pipeline also samples
eight candidates per query and keeps the one with the highest Judge and
CLAP scores, a post-processing step absent from the other methods.
Applying it to \method{} and both SAM~Audio checkpoints leaves a
significant Judge gap, and the speech measures, which the reranker does
not select on, stay far apart (Appendix Table~\ref{tab:k8veo}). The
candidate-count paragraph of Appendix~\ref{sec:candidates} explains why
we treat single draws as the fairer comparison.
\textbf{LTX-2.3 benchmark.} \method{} leads every cascade baseline on
every metric (Appendix Table~\ref{tab:ltx}).
\textbf{SAM-Audio-bench.} On real recordings, scored with a Judge from
the same encoder family as SAM~Audio, \method{} leads the strongest
SAM~Audio checkpoint in Judge and every cascade baseline on both swap
margins (Table~\ref{tab:sambench}). Captions follow our style (e.g.,
``Car driving'' becomes ``A car, driving.''). A comparison with the
benchmark's native text is presented at Appendix Table~\ref{tab:sam172}.

\subsection{User study}
\label{sec:userstudy}

\begin{figure}[tb]
    \centering
    \begin{minipage}[t]{0.485\linewidth}
      \centering
      \includegraphics[trim=0 0.49in 0 0.13in, clip, width=\linewidth]{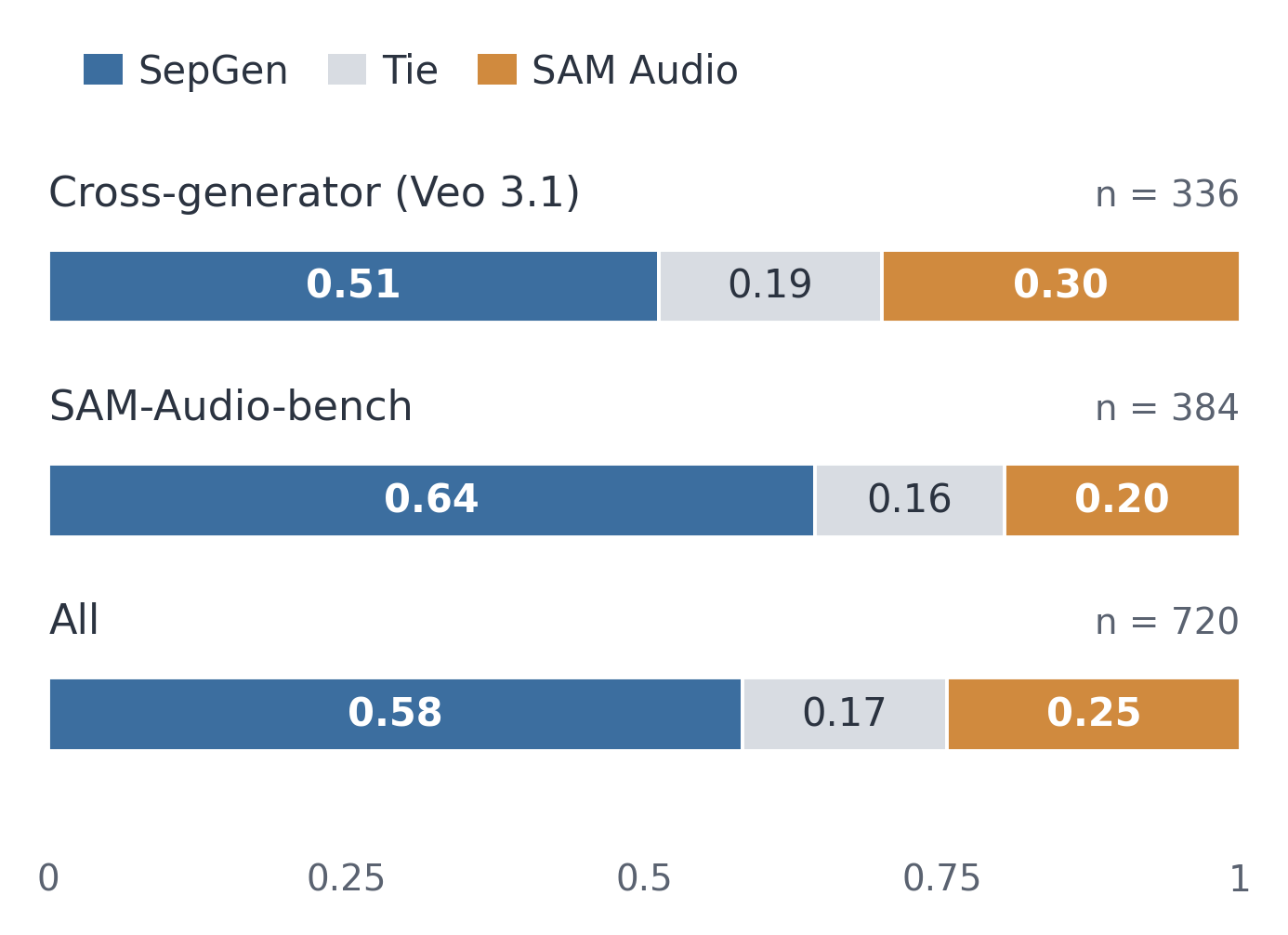}
    \end{minipage}\hfill
    \begin{minipage}[t]{0.485\linewidth}
      \centering
      \includegraphics[trim=0 0.49in 0 0.13in, clip, width=\linewidth]{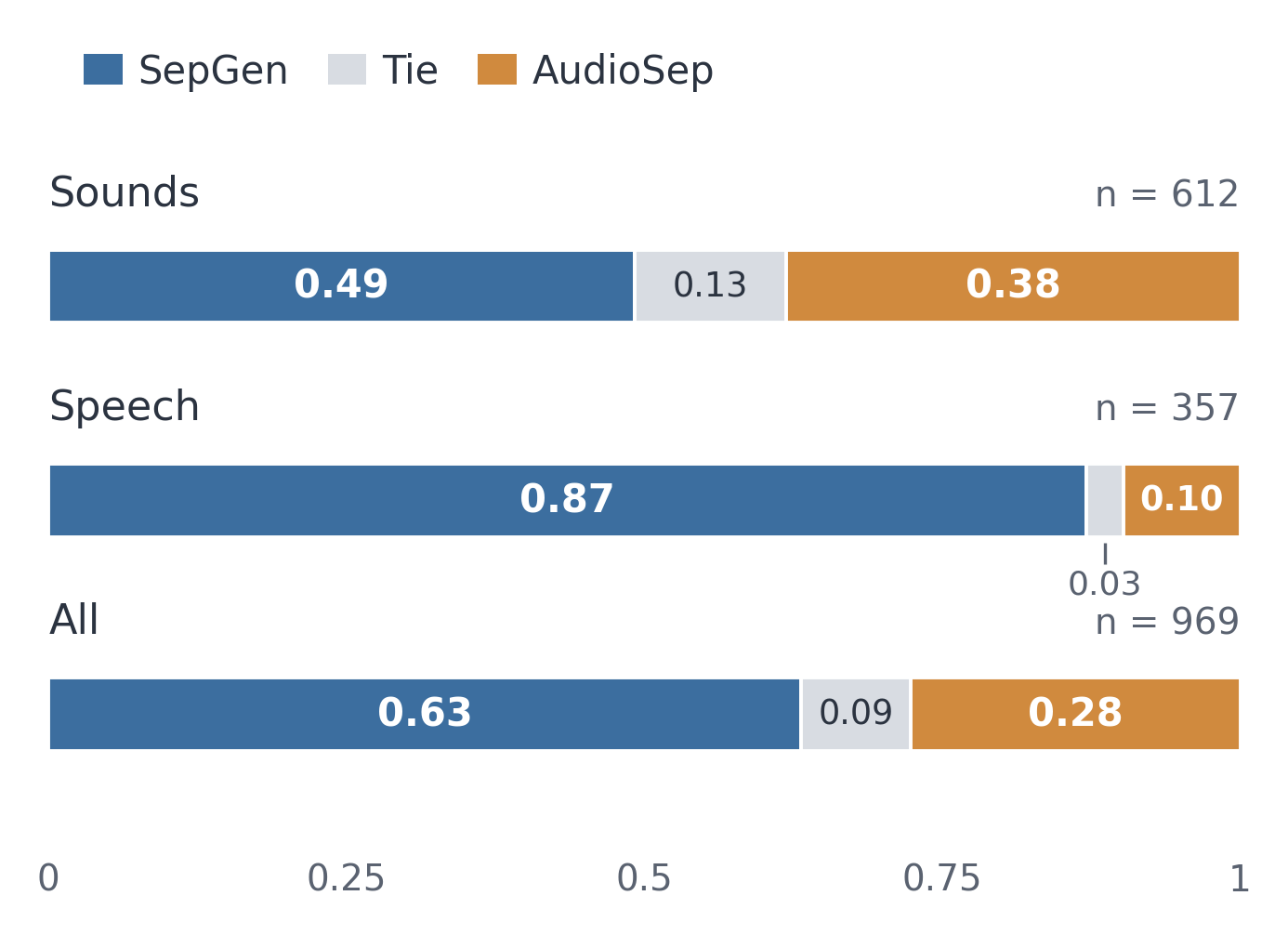}
    \end{minipage}
    \caption{\textbf{User study.} Forced choice with a tie option, against SAM~Audio
    \textbf{(a, left)} on ten clips from Tables~\ref{tab:sep} and~\ref{tab:sambench},
    48 listeners, and AudioSep \textbf{(b, right)} on ten generation clips of
    Table~\ref{tab:gen}, 51 listeners. Bars give the share of the \(n\) judgements.}
    \label{fig:userstudy}
    \vspace{-0.1in}
  \end{figure}

Figure~\ref{fig:userstudy} reports two listening studies, one per mode,
comparing \method{} with a cascade baseline: SAM~Audio
\citep{samaudio2026} for separation and AudioSep
\citep{liu2024audiosep} for generation. For each source caption,
participants viewed the reference video, listened to the two anonymized
candidate stems, in an order randomized per listener and source, and
selected the preferred stem or a tie; the protocol is detailed in
Appendix~\ref{sec:userprotocol}.
The separation study runs SAM~Audio's released pipeline unchanged, so
its ranker selects the best of eight candidates per source
(\(k{=}8\)) while \method{} contributed a single draw. In separation,
\method{} received 0.58 of the judgements and SAM~Audio 0.25 (panel
a), clearly on the real recordings of SAM-Audio-bench (0.64 against
0.20) and by a narrower margin on the Veo benchmark (0.51 against
0.30). In generation, \method{} received 0.63 and AudioSep 0.28 (panel
b), 0.87 against 0.10 on speech and 0.49 against 0.38 on sounds.

\subsection{Ablations}
\label{sec:ablations}

\begin{table}[t]
  \centering
  \caption{\textbf{Ablations} on the LTX-2.3 benchmark (\ckptsep{}, seed 0). Each block modifies a single component of our configuration. Stars mark Holm-corrected paired Wilcoxon significance against our
configuration (*~\(p<0.05\), **~\(p<0.01\), ***~\(p<0.001\)).}
  \label{tab:ablations}
  \small
  \setlength{\tabcolsep}{2.6pt}
  \begin{tabular}{llccccccc}
    \toprule
    & & \multicolumn{5}{c}{Sounds} & \multicolumn{2}{c}{Speech}\\
    \cmidrule(lr){3-7}\cmidrule(lr){8-9}
    Component & Condition & Judge \(\uparrow\) & J-swap \(\uparrow\)
      & CLAP \(\uparrow\) & C-swap \(\uparrow\) & Sib-sim \(\downarrow\)
      & WER \(\downarrow\) & Cosine \(\downarrow\)\\
    \midrule
    \multicolumn{2}{l}{\emph{Our configuration}}
      & 4.07 & \underline{1.99} & \textbf{0.19} & \textbf{0.18} & \textbf{0.01} & \textbf{0.09} & \textbf{0.66}\\
    \midrule
    Jointness & Independent passes
      & 2.43\textsuperscript{***} & 0.21\textsuperscript{***} & \underline{0.16}\textsuperscript{*} & 0.04\textsuperscript{***} & 0.13\textsuperscript{***} & 1.05\textsuperscript{***} & 0.95\textsuperscript{***}\\
    \midrule
    Video & Blank
      & \textbf{4.09} & \textbf{2.01} & \textbf{0.19} & \textbf{0.18} & \textbf{0.01} & \textbf{0.09} & \underline{0.67}\\
     & Unrelated
      & \underline{4.08} & \textbf{2.01} & \textbf{0.19} & \underline{0.17} & \textbf{0.01} & \textbf{0.09} & 0.68\\
     & None
      & 4.06 & \underline{1.99} & \textbf{0.19} & \underline{0.17} & \textbf{0.01} & \underline{0.13} & \underline{0.67}\\
    \midrule
    Aligned Positions & Retrained
      & 3.61\textsuperscript{**} & 1.36\textsuperscript{**} & \textbf{0.19} & 0.14 & \underline{0.05}\textsuperscript{*} & 0.31\textsuperscript{**} & 0.69\\
    \bottomrule
  \end{tabular}
  \vspace{-0.1in}
\end{table}

Table~\ref{tab:ablations} ablates each component on the LTX-2.3 benchmark. Sibling similarity (Sib-sim) is the CLAP score of a stem against its sibling's caption.

\textbf{Jointness.} Separating each caption in a standalone pass, without the
sibling stem, returns two near-copies of the audio-mix and collapses every
swap margin, word error and speaker cosine; generation collapses the same
way (Appendix Table~\ref{tab:genjoint}). The joint pass is what gives the two
stems complementary content.
\textbf{Protected audio-mix.} A generation pass executed with the
adapter enabled and disabled from identical inputs yields bit-identical
audio-mix tokens and video at every layer and at the output; the stem
tokens differ (Appendix~\ref{sec:settings}).
\textbf{Video.} Blank frames, an unrelated clip, or no video leave every
separation metric unchanged (no significant paired difference): each
stem's video attention finds its source (Figure~\ref{fig:stemattn}), but
the captions and audio-mix drive separation.
\textbf{CSAG.} Appendix Table~\ref{tab:csagabl} ablates the guidance strength.
\textbf{Aligned positions.} Retraining with a concatenated rotary
timeline instead of the aligned layout drops Judge and more than
triples WER, indicating that the stems depend on their temporal alignment with the
audio-mix.
\textbf{Audio-mix conditioning.} Appendix Table~\ref{tab:mixablation} shows that
the stems are separated from the audio-mix rather than synthesized.

\section{Discussion and Conclusion}
\label{sec:discussion}

\noindent \textbf{Limitations.} The audio VAE decoder keeps each stem's
envelope and spectrum and regenerates its phase, so SI-SDR against a
reference waveform measures that phase resynthesis
(Appendix~\ref{sec:sisdr}). The current layout is fixed at two
sources. The stems keep the acoustics of the scene they come from, including
reverberation and distance cues, so a spatial renderer must account for
them before propagating each source to a new viewpoint.

\noindent \textbf{Conclusion.} \method{} allows a pretrained audio-video
generator to produce per-source waveforms alongside its video and
soundtrack, by training a lightweight stem adapter on paired
two-source data and switching between separation and generation
through the audio-mix noise schedule alone. The stems are free-text
addressed, jointly denoised, and grounded in a protected audio-mix
that the backbone produces unmodified. Future work includes extending
the layout beyond two stems, and evaluating a placement readout from
each stem's video attention (Appendix~\ref{sec:locapp}) and
integrating it into a spatial renderer for the 4D scenes that motivate
this work. The project page (\url{https://sepgen.github.io/}) shows
audio-video examples of both modes beside the baselines, and an
explainer video.

\bibliographystyle{iclr2027_conference}
\bibliography{refs}

\clearpage
\appendix
\renewcommand{\floatpagefraction}{0.6}
\renewcommand{\bottomfraction}{0.6}
\setcounter{table}{0}\renewcommand{\thetable}{S\arabic{table}}
\setcounter{figure}{0}\renewcommand{\thefigure}{S\arabic{figure}}
\begin{center}{\Large\bfseries Appendix}\end{center}
\makeatletter\setlength{\@fptop}{0pt}\setlength{\@fpsep}{12pt plus 2pt}\setlength{\@fpbot}{0pt plus 1fil}\makeatother

This appendix gives training
and inference details (\S\ref{sec:traindata}), benchmark construction
(\S\ref{sec:evalsets}), the evaluation protocol (\S\ref{sec:metrics}),
full generation and separation results (\S\ref{sec:gentables},
\S\ref{sec:septables}), a waveform-level evaluation against reference
stems (\S\ref{sec:refstem}), and the placement readout
(\S\ref{sec:locapp}). The project page
(\url{https://sepgen.github.io/}) has audio and video examples of both
modes next to the cascade baselines, the main results tables, the
user-study results, and an explainer video. We encourage listening to the examples, since automatic metrics are
proxies for what a listener hears.

\FloatBarrier
\section{Training and inference details}
\label{sec:traindata}

Training uses 4{,}843 two-source segments from four families:
CelebV-HQ talking-head pairs (\(n{=}1{,}658\))
\citep{zhu2022celebvhq}, URMP two-instrument scenes
(\(n{=}1{,}437\)) \citep{li2018urmp}, generated two-source clips
(\(n{=}749\)), and cross-video composites (\(n{=}999\)) of
VGGSound \citep{chen2020vggsound} and MUSIC-21
\citep{zhao2019soundofmotions} clips. Ground-truth stems exist in
every family because each audio-mix is the sum of two known
sources: URMP provides per-instrument tracks, and the other
families mix two single-source clips (two single-speaker
recordings, two clips synthesized by our model, or two web
clips placed as split-screen panels or with one source off screen).
Each source caption comes in three registers, semantic,
positional, and short, sampled with equal probability. Positional
captions state where each source is located, and a \emph{source
box}, the frame region a source occupies in each video latent
frame according to the composition records, enforces that
statement during training through a soft additive bias that
restricts each stem's audio queries in video-to-audio attention to
the video tokens inside its box. The bias is zeroed on the
audio-mix channel and removed at inference, so no source location
is supplied to the model when it separates. Since
SAM-Audio-bench and three training families are YouTube clips, we
verified that no video contributes to both: the 771 VGGSound, 170
MUSIC-21, and 133 CelebV-HQ training videos share no identifier
with the 592 videos of the full SAM-Audio-bench release, and the
benchmark's VGGSound rows resolve in the same index as our
training clips.

\paragraph{Settings.}
\label{sec:settings}
Training runs in two stages, 12{,}000 separation updates followed
by 3{,}000 generation updates, on the audio stream with the
LTX-2.5 22B weights frozen and int8-quantized, on a single NVIDIA
A6000: AdamW, batch size 1, learning rate
\(10^{-4}\) with linear decay, bfloat16, LoRA rank and scale 128
\citep{hu2022lora} on audio self-attention, the query and output
projections of text cross-attention and video-to-audio attention,
and the audio feed-forward block, 480 modules and 327M trainable
parameters in total. Training segments are 113 frames
at 25\,fps encoded at \(768\times 512\). The LoRA delta is zeroed on
audio-mix tokens (the protected audio-mix of the main paper, Eq.~\eqref{eq:lora-gate})
and the loss is computed on the stems alone. The
separation stage presents the audio-mix clean (\(\sigma{=}0\)) at
every step; the generation stage presents it clean on 30\% of
samples and otherwise at an independent noise level clamped to that
of the stems, \(\sigma_m=\min(\sigma'_m,\sigma_s)\). At inference, video is encoded at \(768\times 512\) and each clip
keeps its length (113 frames at 25\,fps on the generation
benchmark; 121, 137, and 241 frames at 24\,fps on LTX-2.3, Veo,
and SAM-Audio-bench). Generation adds an attention bias from stem
queries to audio-mix keys, \(b_0(1-\rho)U+b_fD_w\) with
\(b_0{=}b_f{=}1\) and \(w{=}4\), where \(U\) covers every audio-mix
key, \(D_w\) the keys within \(w\) latent frames of the query's
own time, and \(\rho\) the fraction of the schedule elapsed, so the
uniform term fades while the time-local band remains; once
\(\sigma\le 0.97\), Estimated Separation (Section~\ref{sec:schedule}
of the main paper) conditions the stems on the clean estimate of
the audio-mix. Each generation pass adds a second-stage spatial
upscale and returns \(1536\times 1024\) video at 113 frames and
25\,fps with three 48\,kHz waveforms. Separation re-clamps the encoded
audio-mix to \(\sigma{=}0\) at every step. Guidance follows the
LTX-2 pipeline defaults: classifier-free guidance (CFG) scales of
3.0 for video and 7.0 for audio, and, in generation, cross-modal guidance 3.0
on the video and audio-mix tracks; the stems receive no isolated-modality
guidance in either mode. Separation uses 30 Euler steps with CFG
rescale 0.7; generation uses the LTX pipeline's 15-step
\texttt{res\_2s} sampler with video/audio-mix rescales 0.45/1.0 and
stem rescale 0.7.

\begin{figure*}[t]
  \centering
  \includegraphics[width=\linewidth]{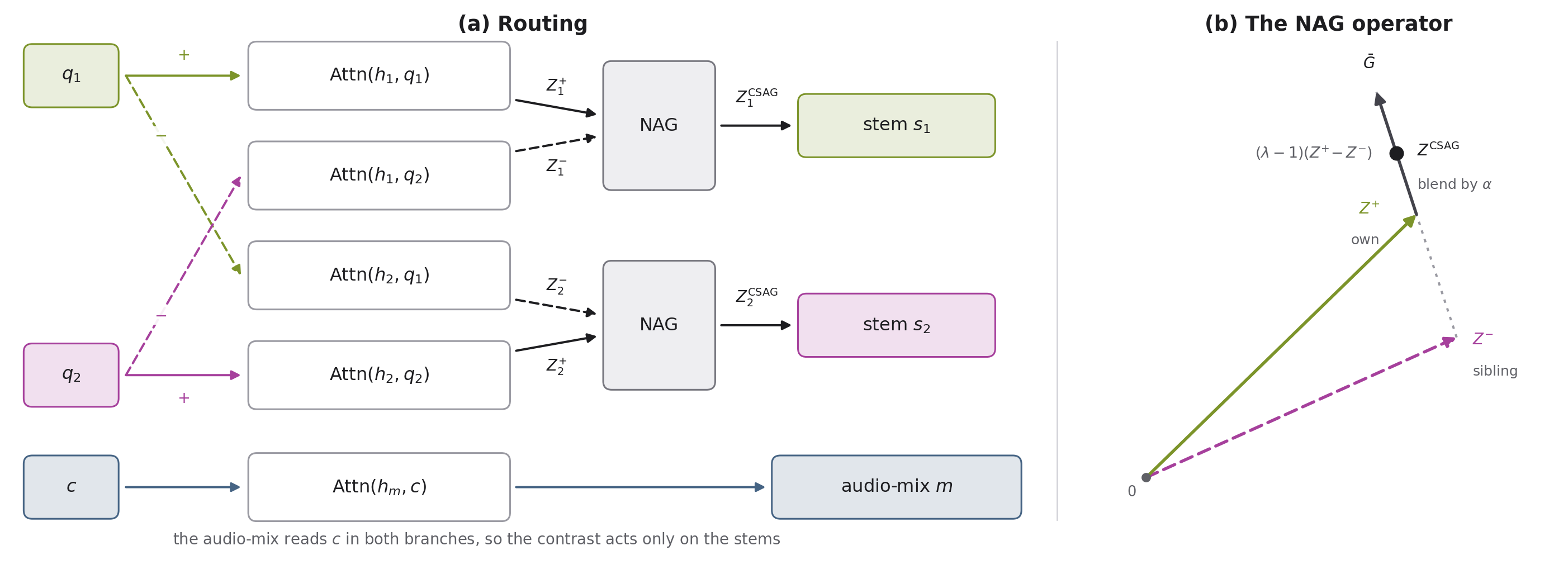}
  \caption{\textbf{Cross-Stem Attention Guidance.}
  (a)~Each stem attends to its own caption in the positive branch (solid) and
  its sibling's in the negative branch (dashed). The audio-mix attends to
  \(c\) in both.
  (b)~NAG moves \(Z^{+}\) away from \(Z^{-}\) to obtain
  \(Z^{\mathrm{CSAG}}\).}
  \label{fig:csag2}
\end{figure*}

\paragraph{Cross-Stem Attention Guidance.}
\label{sec:csag}
Routing each stem to its own caption does not ensure that each stem
contains only its own source: when a sound fits both captions, both
stems may reproduce it. Cross-Stem Attention Guidance (CSAG) steers
each stem away from its sibling's caption inside the caption
cross-attention. For the hidden states \(h_i\) of stem \(i\), CSAG
computes a positive branch with the stem's own caption \(q_i\) and a
negative branch with the sibling caption \(q_j\)
(Figure~\ref{fig:csag2}a),
\begin{equation}
  \begin{aligned}
  Z_i^+&=\operatorname{Attn}(h_i,q_i),&
  Z_i^-&=\operatorname{Attn}(h_i,q_j),\\
  Z_i^{\mathrm{CSAG}}&=\operatorname{NAG}(Z_i^+,Z_i^-),&
  &j\neq i ,
  \end{aligned}
  \label{eq:nag}
\end{equation}
where \(Z\) denotes an attention output. The branches are combined by
the normalized attention guidance operator of \citet{nag2025}
(Figure~\ref{fig:csag2}b),
\begin{equation}
  \begin{aligned}
  G &= \lambda Z^+-(\lambda-1)Z^-,\\
  r &= \frac{\lVert G\rVert_1}{\lVert Z^+\rVert_1},\qquad
  \bar G =G\,\frac{\min(r,\tau)}{r},\\
  Z^{\mathrm{CSAG}}&=\alpha\bar G+(1-\alpha)Z^+ ,
  \end{aligned}
\end{equation}
with per-token norms over the feature dimension; \(\lambda\)
corresponds to an extrapolation coefficient of \(\lambda-1\) in the
notation of \citet{nag2025}, \(\tau\) bounds the norm ratio, and
\(\alpha\) blends the result with the positive branch. We use
\(\lambda{=}2\), \(\tau{=}2.5\), and \(\alpha{=}0.5\), in every
audio-stream caption cross-attention block at every step. The
audio-mix reads the scene caption in both branches and so carries no
contrast. On the LTX-2.3 benchmark, removing the guidance lowers the
CLAP swap margin (Holm \(p{=}0.03\)); Judge and J-swap move in the
same direction without reaching significance, and \(\lambda{\ge}3\)
lowers the Judge (Table~\ref{tab:csagabl}).

\begin{table}[htbp]
  \centering
  \caption{Cross-Stem Attention Guidance ablation on the LTX-2.3 benchmark (\ckptsep{}, seed 0); stars mark Holm-corrected paired Wilcoxon significance against \(\lambda{=}2\), as in Table~\ref{tab:ablations} of the main paper.}
  \label{tab:csagabl}
  \small
  \setlength{\tabcolsep}{3pt}
  \begin{tabular}{lccccccc}
    \toprule
    & \multicolumn{5}{c}{Sounds} & \multicolumn{2}{c}{Speech}\\
    \cmidrule(lr){2-6}\cmidrule(lr){7-8}
    Guidance & Judge \(\uparrow\) & J-swap \(\uparrow\) & CLAP \(\uparrow\) & C-swap \(\uparrow\) & Sib-sim \(\downarrow\) & WER \(\downarrow\) & Cosine \(\downarrow\)\\
    \midrule
    \(\lambda{=}2\) (ours) & \textbf{4.07} & \textbf{1.99} & \textbf{0.19} & \textbf{0.18} & \textbf{0.01} & \underline{0.09} & \textbf{0.66}\\
    Off & 3.92 & \underline{1.80} & \underline{0.18} & 0.15\textsuperscript{*} & \underline{0.03} & \textbf{0.08} & \underline{0.68}\\
    \(\lambda{=}3\) & \underline{3.94}\textsuperscript{*} & 1.77\textsuperscript{*} & \underline{0.18} & \textbf{0.18} & \textbf{0.01} & 0.16 & \textbf{0.66}\\
    \(\lambda{=}5\) & 3.87\textsuperscript{*} & 1.74 & \underline{0.18} & \underline{0.16} & \textbf{0.01} & 0.17 & \textbf{0.66}\\
    \bottomrule
  \end{tabular}
\end{table}

\paragraph{Adapter contribution to the audio-mix and video.}
A denoising pass of the generation configuration is executed twice
from identical inputs, with the adapter enabled and with all LoRA
updates disabled, and the intermediate activations of each
transformer block and the final outputs are compared. The audio-mix
rows and the video are bit-identical between the two runs at every
layer; the stem rows differ. Relative to a stock render from
the same inputs, the audio-mix and video differ only through
floating-point rounding, which arises from the text encoder's batch
composition, the batched guidance layout of the stock pipeline, and
the timestep embedding computed over the three-track sequence, and,
over a full render, through the sampler's noise draw.

\paragraph{Checkpoints.}
\label{sec:repro}
Training yields two checkpoints. The separation checkpoint, taken
after the first stage, is trained on separation only. The
generation checkpoint, taken after the second stage, performs
joint video-and-stem generation and separation with the same
weights and produces the generation results; on the separation
benchmarks, results name their checkpoint, and unnamed separation
results are those of the separation checkpoint.

\FloatBarrier
\section{Benchmarks}
\label{sec:evalsets}

Evaluation spans four benchmarks, three of generated videos, scored
separately for sounds and speech, and one of real recordings,
scored as one pool. On the generated benchmarks, speech captions
consist of a role and a quoted line. Every language-queried method receives identical
per-source captions \citep{samaudio2026,flowsep2025,liu2024audiosep};
SA\(\times\)SAM3 is SAM large-tv driven by SAM3 video masks, and
its empty-mask stems (24 of the 298 Veo stems, 14 of the 74
LTX-2.3 sounds stems) are silent and remain in every mean.

\paragraph{Generation benchmark.}
\label{sec:genbench}
The generation benchmark comprises 194 two-source scene prompts:
speech (\(n{=}60\)) and foley (59), music (45), and speech-with-background
(30) as the sounds lane (\(n{=}134\)). Prompts were drafted with a
large language model and approved manually, and no prompt string
matches a training caption exactly or at a token-set Jaccard
similarity of 0.8 or above. For each prompt, the generation
checkpoint synthesizes the video, the audio-mix, and both stems in
one sampling run; every cascade baseline then receives that
audio-mix with the same two captions, and self-separation passes
it through our separation pipeline. Scorer queries are fixed
acoustic descriptions of each source, shared verbatim with the Veo
benchmark. The generated audio-mixes carry the scripted lines:
transcribed with the speech scorer's recognizer, they read word
error 0.05 against the two lines concatenated. Each clip lasts
4.5\,s.

\paragraph{LTX-2.3 and Veo benchmarks.}
The LTX-2.3 benchmark comprises 64 two-source clips of 5\,s (37
sounds, 27 speech) whose audio-mixes were rendered by LTX-2.3, a
model of the same family as our backbone; captions and scene names
are disjoint from the training segments. The Veo benchmark was rendered by
Veo~3.1 Lite from the same prompt suite as the generation
benchmark with an ultrawide camera cue. Before any separator was
scored, one reviewer screened all 198 renders for a usable
two-source scene, both captioned sources present in picture and
soundtrack, and retained 153; removing four same-prompt duplicates
leaves 149 clips, foley (39), music (36), and
speech-with-background (29) as the sounds lane (\(n{=}104\)) and
turn-taking and overlapping speech as the speech lane
(\(n{=}45\)). Veo renders the scripted lines: over all 90 quoted
lines of the speech lane, the transcript of the audio-mix has a mean
word error of 0.02 against each line. Native clips are \(1280\times 720\) at 144 frames (6\,s)
with stereo 48\,kHz audio; every method is scored on the same first 5.7\,s,
adjusted to the input-length quantization of \method{}.

\paragraph{SAM-Audio-bench.}
SAM-Audio-bench consists of in-the-wild YouTube recordings from
AudioSet, AVSpeech, Condensed Movies, music duets, and VGGSound,
each with one target description, a benchmark-provided 10\,s
window, and no reference stems. We score
its instrument-wild, sound-effects, and speaker subsets, removing
seven speech rows that duplicate another row's video, window, and
text, for 172 recordings (69 / 55 / 48). Every method is queried with
the benchmark's text adapted to our caption style; the cascade
baselines are also scored with the native text
(Table~\ref{tab:sam172}). \method{} writes two
stems, the second
describing the residual background; Judge and CLAP are reported on
the queried target, and the swap-margin sibling text is drawn from
the benchmark's list of other sources present in the clip.

\FloatBarrier
\section{Evaluation protocol}
\label{sec:metrics}

Let \(J(s, c)\) denote the SAM~Audio Judge overall rating of stem
\(s\) given description \(c\) and the input audio-mix, and
\(C(s, c)\) the CLAP cosine between stem and description
\citep{samaudio2026,wu2023laionclap}. For a clip with stems
\(s_1, s_2\) and descriptions \(c_1, c_2\), identity scores are
\(J(s_i, c_i)\) and \(C(s_i, c_i)\), and swap margins are
\(J(s_i, c_i) - J(s_i, c_j)\) and \(C(s_i, c_i) - C(s_i, c_j)\)
for \(j \ne i\), averaged over both stems (J-swap, C-swap); a stem
that copies the audio-mix obtains a margin of zero. Speech stems
are scored by own-line word error, the word error rate of each
stem's transcript against its own speaker's quoted line
\citep{radford2023whisper}, and by speaker cosine between the
speaker embeddings of the two stems, which equals 1.00 for a
duplicated stem. Lower cosine means more distinct voices; a silent stem
also lowers it but scores an own-line word error near 1.0, so the
two are read together. CLAP scores use LAION-CLAP with the HTSAT-base
audio encoder and the
\texttt{music\_speech\_audioset\_epoch\_15\_esc\_89.98} checkpoint;
transcripts come from Whisper large-v3 (faster-whisper); speaker
embeddings are the x-vectors of
\texttt{microsoft/wavlm-base-plus-sv}. Judge shares the PE-AV (Perception Encoder, audio-visual)
encoder family with the SAM~Audio separator, and AudioSep's query
encoder is the scorer's CLAP checkpoint, which inflates its
CLAP-derived columns. CLAP identity places the audio-mix returned
twice close to a real method (0.12 on the sounds lane, against
0.07 for silence), whereas swap margin places both controls at
0.00; the headline claims therefore rest on margins. The
reference-stem evaluation of Section~\ref{sec:refstem} adds two
waveform-level measures.

\paragraph{Seeds and statistics.}
\label{sec:seeds}
Separation tables report mean \(\pm\) standard deviation over
seeds \(\{0,1,2\}\), and SAM-Audio-bench over \(\{0,\dots,4\}\); a
seed redraws each method against the same fixed audio-mix, and
AudioSep, being deterministic, enters as one draw, and its
word-error spread reflects recognizer variation alone. On the
generation benchmark a seed re-runs the entire experiment, so the
spread also carries the variation of the generated audio-mix.
Every
\(p\) is a Holm-corrected paired Wilcoxon test within its stated
family. Whisper transcription is
not deterministic (pass-to-pass spread up to 0.08 own-line word
error), so word-error comparisons are made within one recognizer
pass. Tables report mean \(\pm\) SD over these seeds, best in bold and
second underlined; stars on a baseline cell give the significance of
\method{} against it at the weakest seed: *** \(p<0.001\),
** \(p<0.01\), * \(p<0.05\), ns not significant.

\paragraph{Metric controls.}
\label{sec:metriccontrols}
Three controls bound alternative readings of the scores. Our stems
come out of our audio VAE decoder. Each cascade baseline uses its own
decoder or mask, so a scorer could favor the sound of our decoder. We
pass each baseline's stems through our audio VAE and rescore them
(Table~\ref{tab:requant}). The ranking does not change, and the
decoder accounts for at most 27\% of a baseline's Judge margin (0.15
of AudioSep's 0.55).
FlowSep delivers 16\,kHz audio and AudioSep 32\,kHz; band-limiting
every method's stems and the scorer's references to 16\,kHz keeps
96\% of the Judge gap, 94\% of J-swap, and 76\% of C-swap over the
best cascade baseline, and \method{} stays ahead of every cascade
baseline. The summed stems
carry the spectral content of the audio-mix without matching its
samples (Section~\ref{sec:sisdr}), and the WavLM speaker
embedding, outside the Judge family, confirms that each speech
stem carries the voice of its audio-mix: median stem-to-audio-mix
cosine is 0.97 to 0.98, against 0.60 to 0.70 for wrong-voice
anchors, and the own-region minus sibling-region margin is
\(+0.21\) to \(+0.33\), against at most \(+0.04\) for every cascade
baseline.

\begin{table}[htbp]
  \centering
  \caption{Audio VAE requantization control (sounds, \(n{=}134\), seed 0). The stems of each method are passed once more through the audio VAE; \(\Delta\) denotes the requantized minus the published score.}
  \small
  \begin{tabular}{lcc}
    \toprule
    Method & Judge (published) & \(\Delta\)\\
    \midrule
    FlowSep & 3.49 & \(+0.08\)\\
    AudioSep & 3.27 & \(+0.15\)\\
    SAM large-tv & 2.84 & \(-0.09\)\\
    SAM large & 2.75 & \(-0.08\)\\
    \midrule
    \method{} (second round trip) & 3.82 & 0.00\\
    \bottomrule
  \end{tabular}
  \label{tab:requant}
\end{table}

\paragraph{User study protocol.}
\label{sec:userprotocol}
Each study uses ten clips, each with one caption per source to be recovered, and the listener answers one question per caption. The separation study draws four clips from the Veo benchmark and six from SAM-Audio-bench, and the generation study ten clips of the generation benchmark at seed~0. For separation, \method{} contributes one draw per clip at a randomly drawn seed, and SAM~Audio its best-of-eight output, reranked by its own ranker: SAM~Audio large with our captions on the Veo clips, and SAM~Audio large-tv with the native captions on SAM-Audio-bench. For every clip, the listener watches the reference video with its unseparated soundtrack, then hears the two candidate stems of a caption, one per method, and picks the better one, with a tie option. The question asks which candidate is better overall, judged on faithfulness, separation, and alignment with the text description, all three explained on the instruction page. The candidates are anonymous, and the order of the clips and of the two players is drawn independently for every listener and every source, so nothing on the page reveals a method. One source per study is a screening question whose better candidate is obvious, which detects listeners who answer without listening; a listener who misses it is discarded in full, and that source is excluded from the figure, since the remaining listeners all answered it identically. We recruited on the Prolific platform and kept 48 listeners for separation, after two missed the screening question, and 51 for generation. The ten separation clips carry 16 sources, one the screening question, giving \(48\times15=720\) judgements, and the ten generation clips carry two stems each, one of them the screening question, giving \(51\times(10\times2-1)=969\)
judgements. Participants consented to taking part, gave no personal
data beyond an anonymous platform identifier, and were paid a fixed
fee.

\paragraph{User study uncertainty.}
Table~\ref{tab:userstudy_ci} gives each bar of the main paper's user-study figure with a 95\% interval from a bootstrap that resamples listeners and clips independently (10{,}000 draws). The width of the intervals comes mostly from the clips: resampling listeners alone gives 0.55 to 0.61 for the overall separation share of \method{}, and resampling clips alone gives 0.48 to 0.70.

\begin{table}[t]
  \centering
  \small
  \caption{User-study preference shares with 95\% bootstrap intervals, resampling listeners and clips. The Sources column reports the number of sources on which \method{} won, tied, or lost by majority vote.}
  \label{tab:userstudy_ci}
  \begin{tabular}{llrlclc}
    \toprule
    Study & Cut & \(n\) & \method{} & Tie & Baseline & Sources \\
    \midrule
    Separation & Veo & 336 & 0.51 [0.40, 0.58] & 0.19 & 0.30 [0.19, 0.42] & 3/0/4 \\
     & SAM-Audio-bench & 384 & 0.64 [0.45, 0.83] & 0.16 & 0.20 [0.09, 0.34] & 6/0/2 \\
     & All & 720 & 0.58 [0.47, 0.70] & 0.17 & 0.25 [0.16, 0.34] & 9/0/6 \\
    \midrule
    Generation & Sounds & 612 & 0.49 [0.35, 0.63] & 0.13 & 0.38 [0.24, 0.51] & 6/1/5 \\
     & Speech & 357 & 0.87 [0.80, 0.95] & 0.03 & 0.10 [0.04, 0.17] & 7/0/0 \\
     & All & 969 & 0.63 [0.48, 0.78] & 0.09 & 0.28 [0.15, 0.40] & 13/1/5 \\
    \bottomrule
  \end{tabular}
\end{table}

\FloatBarrier
\section{Generation results}
\label{sec:gentables}

\begin{figure*}[t]
  \centering
  \setlength{\tabcolsep}{1.5pt}
  \renewcommand{\arraystretch}{0.6}
  \begin{tabular}{@{}cccc@{}}
    {\small\bfseries generated video}
      & {\small\bfseries audio-mix}
      & {\small\bfseries source 1}
      & {\small\bfseries source 2}\\[2pt]
    \includegraphics[width=0.245\linewidth,height=0.060\textheight,keepaspectratio]{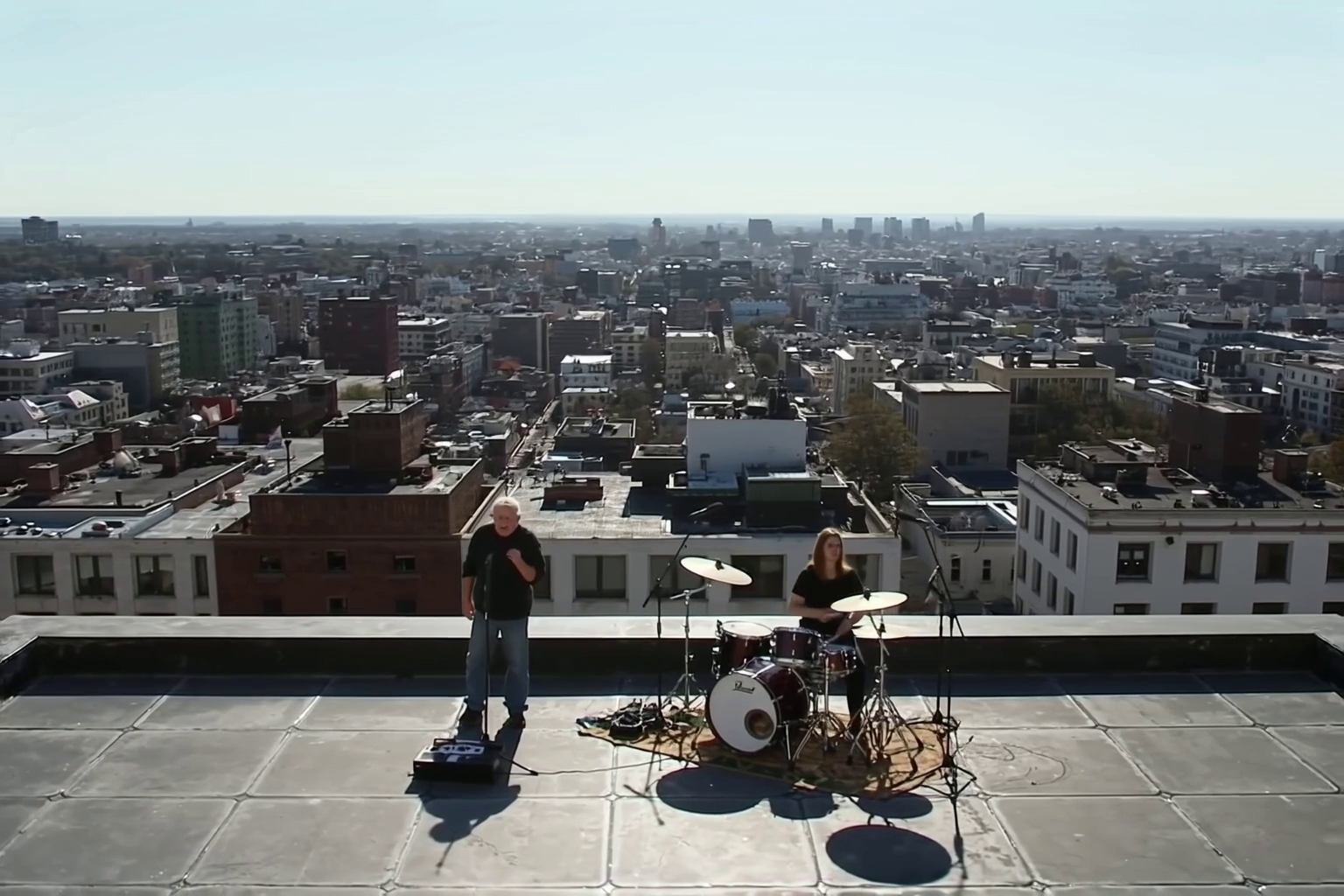}
      & \includegraphics[width=0.245\linewidth,height=0.060\textheight,keepaspectratio]{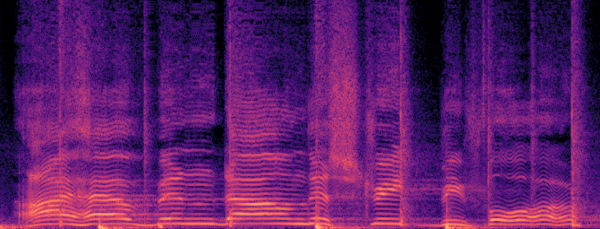}
      & \includegraphics[width=0.245\linewidth,height=0.060\textheight,keepaspectratio]{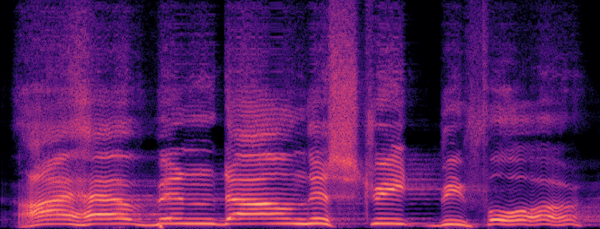}
      & \includegraphics[width=0.245\linewidth,height=0.060\textheight,keepaspectratio]{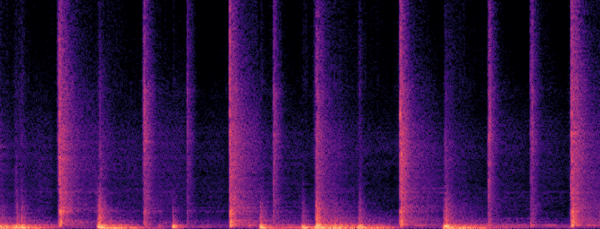}\\[1.5pt]
    \includegraphics[width=0.245\linewidth,height=0.060\textheight,keepaspectratio]{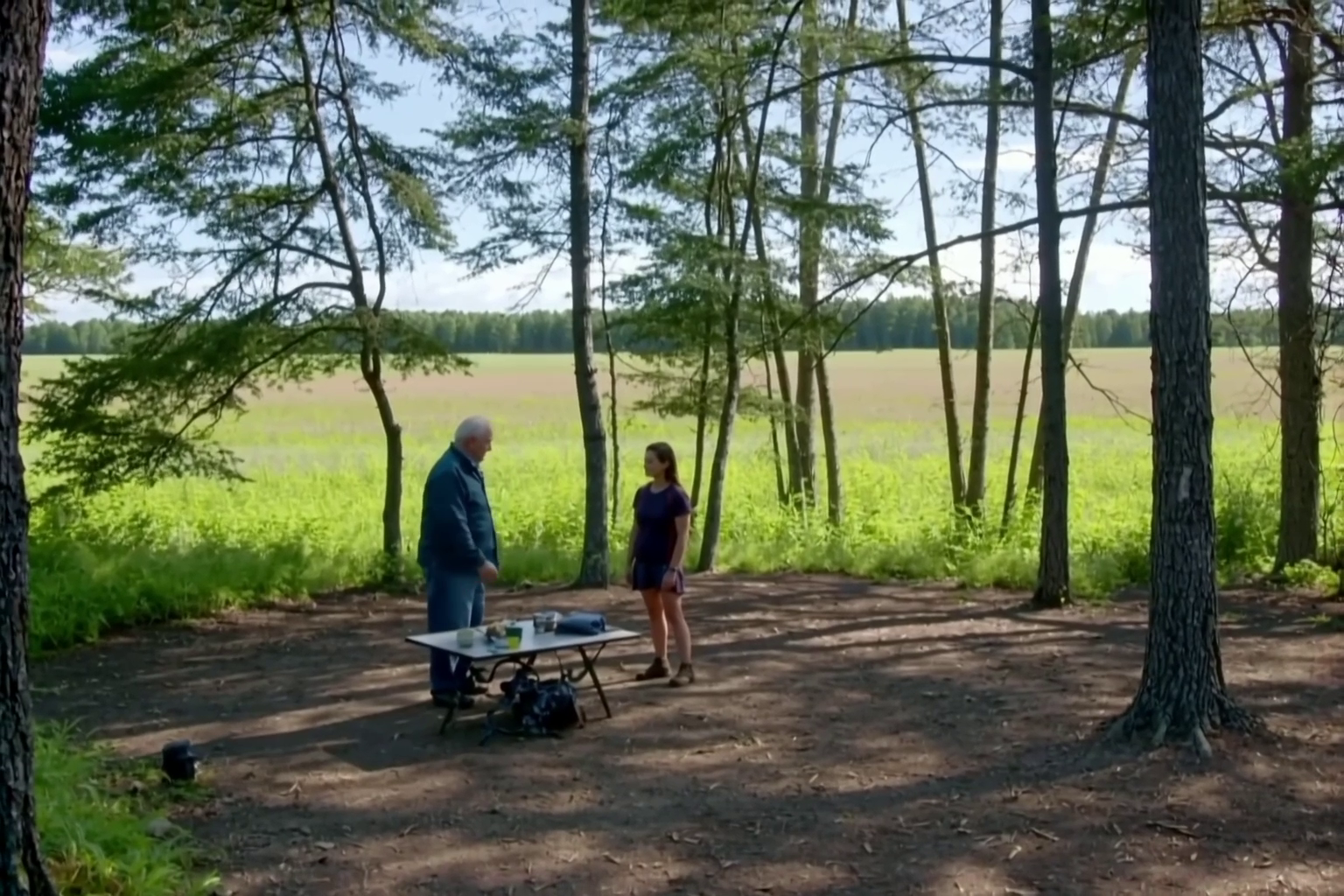}
      & \includegraphics[width=0.245\linewidth,height=0.060\textheight,keepaspectratio]{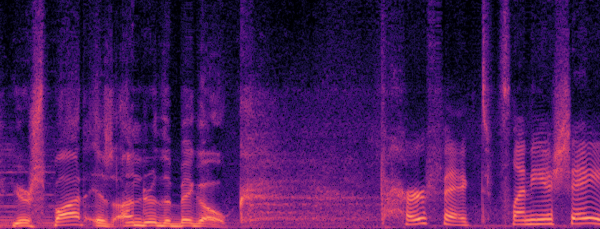}
      & \includegraphics[width=0.245\linewidth,height=0.060\textheight,keepaspectratio]{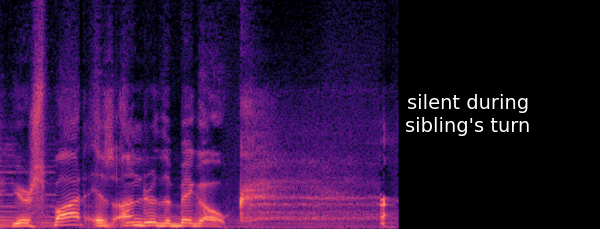}
      & \includegraphics[width=0.245\linewidth,height=0.060\textheight,keepaspectratio]{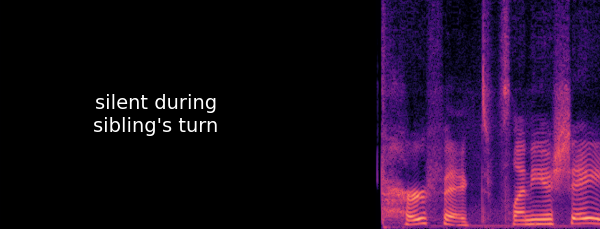}\\
  \end{tabular}
  \caption{\textbf{Generated stems.}
    Top: singing and drums. Bottom: two turn-taking speakers.
    Each row shows the generated video, audio-mix, and caption-addressed
    stems from one sampling run.}
  \label{fig:qualgen}
\end{figure*}

Figure~\ref{fig:qualgen} shows stems generated by \method{} on scenes
of the generation benchmark.

Table~\ref{tab:gen3seed} repeats Table~\ref{tab:gen} of the main paper with
seed spreads, paired-test stars, and three degenerate controls, and
Table~\ref{tab:genblocks} splits the sounds lane by scene category:
the Judge and swap-margin leads hold in every category and are
widest on music, where two
instruments sound for the whole clip and a cascade has no silent
span to exploit. Seed spreads are larger for the cascade baselines
because a cascade's seed mean
moves with the regenerated audio-mixes it is handed (FlowSep reads
3.49, 3.43, and 3.14 Judge on the three seeds), whereas stems generated jointly with the
audio-mix do not inherit its difficulty.

\begin{table}[!htb]
  \centering
  \caption{Generation benchmark with seed standard deviations, degenerate controls, and the paired-test significance of \method{} (joint run) against each baseline.}
  \label{tab:gen3seed}
  \scriptsize
  \setlength{\tabcolsep}{2.2pt}
  \fitwidth{%
  \begin{tabular}{l rrrr rr}
    \toprule
    & \multicolumn{4}{c}{Sounds (\(n{=}134\))} & \multicolumn{2}{c}{Speech (\(n{=}60\))}\\
    \cmidrule(lr){2-5}\cmidrule(lr){6-7}
        Method & Judge \(\uparrow\) & J-swap \(\uparrow\) & CLAP \(\uparrow\) & C-swap \(\uparrow\) & WER \(\downarrow\) & Cosine \(\downarrow\)\\
    \midrule
        FlowSep & \(3.35\pm0.19^{**}\) & \(1.26\pm0.14^{***}\) & \(0.14\pm0.00^{\mathrm{ns}}\) & \(0.10\pm0.01^{***}\) & \(1.03\pm0.04^{***}\) & \(0.97\pm0.01^{***}\)\\
        AudioSep & \(3.23\pm0.05^{***}\) & \(1.27\pm0.04^{***}\) & \(0.14\pm0.01^{\mathrm{ns}}\) & \(0.11\pm0.01^{***}\) & \(0.91\pm0.02^{***}\) & \(0.87\pm0.01^{***}\)\\
        SAM large-tv & \(2.96\pm0.13^{***}\) & \(0.41\pm0.03^{***}\) & \(0.05\pm0.01^{***}\) & \(0.03\pm0.01^{***}\) & \(0.87\pm0.02^{***}\) & \(0.95\pm0.03^{***}\)\\
        SAM large & \(2.93\pm0.19^{***}\) & \(0.42\pm0.11^{***}\) & \(0.04\pm0.01^{***}\) & \(0.03\pm0.01^{***}\) & \(0.81\pm0.06^{***}\) & \(0.95\pm0.02^{***}\)\\
    \midrule
        Audio-mix twice & \(1.54\pm0.04\) & 0.00 & \(0.12\pm0.00\) & 0.00 & \(1.09\pm0.02\) & 1.00\\
        Half silent & \(1.28\pm0.01\) & \(-0.02\pm0.04\) & \(0.10\pm0.00\) & \(0.02\pm0.00\) & \(1.01\pm0.01\) & \(0.48\pm0.01\)\\
        Silent stem & \(1.04\pm0.00\) & 0.00 & 0.07 & 0.00 & 1.00 & 1.00\\
    \midrule
        \method{} (joint run) & \(\underline{3.82\pm0.01}\) & \(\underline{2.07\pm0.01}\) & \(\underline{0.15\pm0.00}\) & \(\underline{0.18\pm0.00}\) & \(\mathbf{0.07\pm0.03}\) & \(\mathbf{0.66\pm0.01}\)\\
        \method{} (self-separation) & \(\mathbf{4.21\pm0.01}\) & \(\mathbf{2.29\pm0.02}\) & \(\mathbf{0.16\pm0.00}\) & \(\mathbf{0.20\pm0.00}\) & \(\underline{0.10\pm0.06}\) & \(\underline{0.67\pm0.01}\)\\
    \bottomrule
  \end{tabular}%
  }
\end{table}

\begin{table*}[!htb]
  \centering
\caption{Generation benchmark by scene category; rankings are computed within each category.}
  \label{tab:genblocks}
  \footnotesize
  \setlength{\tabcolsep}{3.2pt}
  \fitwidth{%
  \begin{tabular}{l rrrr}
    \toprule
        Method & Judge \(\uparrow\) & J-swap \(\uparrow\) & CLAP \(\uparrow\) & C-swap \(\uparrow\)\\
    \midrule
\multicolumn{5}{l}{\emph{Foley (\(n{=}59\))}}\\
        FlowSep & \(3.30\pm 0.18\) & \(1.07\pm 0.17\) & \(0.15\pm 0.01\) & \(0.11\pm 0.01\)\\
        AudioSep & \(3.23\pm 0.10\) & \(1.21\pm 0.03\) & \(\underline{0.16\pm 0.01}\) & \(0.13\pm 0.00\)\\
        SAM large-tv & \(3.22\pm 0.09\) & \(0.38\pm 0.12\) & \(0.07\pm 0.01\) & \(0.03\pm 0.01\)\\
        SAM large & \(3.25\pm 0.19\) & \(0.45\pm 0.17\) & \(0.06\pm 0.01\) & \(0.04\pm 0.01\)\\
    \cmidrule(lr){1-5}
        \method{} (joint run) & \(\underline{3.57\pm 0.01}\) & \(\underline{1.61\pm 0.07}\) & \(\underline{0.16\pm 0.00}\) & \(\underline{0.18\pm 0.00}\)\\
        \method{} (self-separation) & \(\mathbf{4.04\pm 0.01}\) & \(\mathbf{1.81\pm 0.04}\) & \(\mathbf{0.17\pm 0.00}\) & \(\mathbf{0.20\pm 0.00}\)\\
    \midrule
    \multicolumn{5}{l}{\emph{Music (\(n{=}45\))}}\\
        FlowSep & \(3.15\pm 0.19\) & \(0.74\pm 0.09\) & \(0.12\pm 0.00\) & \(0.07\pm 0.01\)\\
        AudioSep & \(2.94\pm 0.02\) & \(0.73\pm 0.07\) & \(\underline{0.13\pm 0.00}\) & \(0.08\pm 0.01\)\\
        SAM large-tv & \(3.01\pm 0.27\) & \(0.49\pm 0.03\) & \(0.06\pm 0.01\) & \(0.04\pm 0.01\)\\
        SAM large & \(2.71\pm 0.31\) & \(0.49\pm 0.21\) & \(0.03\pm 0.03\) & \(0.04\pm 0.02\)\\
    \cmidrule(lr){1-5}
        \method{} (joint run) & \(\underline{3.86\pm 0.01}\) & \(\underline{1.95\pm 0.05}\) & \(\mathbf{0.17\pm 0.00}\) & \(\mathbf{0.19\pm 0.00}\)\\
        \method{} (self-separation) & \(\mathbf{4.28\pm 0.05}\) & \(\mathbf{2.17\pm 0.12}\) & \(\mathbf{0.17\pm 0.00}\) & \(\underline{0.18\pm 0.00}\)\\
    \midrule
    \multicolumn{5}{l}{\emph{Speech-with-background (\(n{=}30\))}}\\
        FlowSep & \(3.76\pm 0.23\) & \(2.40\pm 0.33\) & \(\mathbf{0.13\pm 0.01}\) & \(0.14\pm 0.01\)\\
        AudioSep & \(3.63\pm 0.05\) & \(2.22\pm 0.05\) & \(\underline{0.10\pm 0.01}\) & \(0.14\pm 0.01\)\\
        SAM large-tv & \(2.39\pm 0.27\) & \(0.37\pm 0.11\) & \(-0.02\pm 0.03\) & \(0.02\pm 0.02\)\\
        SAM large & \(2.61\pm 0.27\) & \(0.26\pm 0.11\) & \(0.00\pm 0.02\) & \(0.01\pm 0.01\)\\
    \cmidrule(lr){1-5}
        \method{} (joint run) & \(\underline{4.25\pm 0.03}\) & \(\underline{3.14\pm 0.06}\) & \(\underline{0.10\pm 0.00}\) & \(\underline{0.18\pm 0.01}\)\\
        \method{} (self-separation) & \(\mathbf{4.45\pm 0.03}\) & \(\mathbf{3.41\pm 0.03}\) & \(\mathbf{0.13\pm 0.00}\) & \(\mathbf{0.21\pm 0.01}\)\\
    \bottomrule
  \end{tabular}%
  }
\end{table*}

Generating each source in
its single-caption pass, without a shared audio-mix channel,
collapses both lanes (Table~\ref{tab:genjoint}): nothing binds the
two scenes, so Judge falls below every cascade baseline that
receives the joint audio-mix; the within-architecture cascade on
LTX-2.3 is in Table~\ref{tab:ablations} of the main paper.

\begin{table}[!htb]
  \centering
  \caption{Effect of joint sampling on the generation benchmark (seed 0): a single joint run compared with one single-caption run per source.}
  \small
  \setlength{\tabcolsep}{4pt}
  \begin{tabular}{lcccc}
    \toprule
    & \multicolumn{2}{c}{Sounds} & \multicolumn{2}{c}{Speech}\\
    \cmidrule(lr){2-3}\cmidrule(lr){4-5}
        Method & Judge \(\uparrow\) & J-swap \(\uparrow\) & WER \(\downarrow\) & Cosine \(\downarrow\)\\
    \midrule
        \method{} (joint run) & \textbf{3.82} & \(\mathbf{2.08}\) & \textbf{0.11} & \textbf{0.67}\\
        Independent passes & 2.27 & 0.65 & \underline{0.69} & \underline{0.86}\\
    \midrule
        FlowSep on the joint audio-mix & \underline{3.49} & \underline{1.39} & 1.02 & 0.96\\
    \bottomrule
  \end{tabular}
  \label{tab:genjoint}
\end{table}

\paragraph{Dual mode and schedule.}
\label{sec:twockpt}
Table~\ref{tab:schedule} scores the estimated-separation schedule
of the generation stage (Section~\ref{sec:schedule} of the main
paper). Starting
Estimated Separation at every step (\(\sigma_g{=}\sigma_{\max}\))
rather than below \(\sigma_g{=}0.97\) lowers J-swap at equal Judge and word
error, because the stems then read a clean estimate of an
audio-mix that has not yet committed to its sources; using it at no
step (\(\sigma_g{=}0\)) lowers J-swap and raises word error
while Judge rises, so the schedule purchases
isolation of each speaker's line rather than sound quality. Run as
a generator, the separation checkpoint matches the generation
checkpoint on Judge but trails on J-swap and word error; conversely, the
generation checkpoint separates the Veo benchmark at parity with
the separation checkpoint (Table~\ref{tab:veo}).

\begin{table}[!htb]
  \centering
  \caption{Estimated Separation schedule on the generation benchmark. The threshold \(\sigma_g\) is the noise level below which the stems condition on the clean estimate of the audio-mix rather than on its noisy state.}
  \label{tab:schedule}
  \small
  \setlength{\tabcolsep}{3pt}
  \fitwidth{%
  \begin{tabular}{ll rrrr rr}
    \toprule
    & & \multicolumn{4}{c}{Sounds (\(n{=}134\))} & \multicolumn{2}{c}{Speech (\(n{=}60\))}\\
    \cmidrule(lr){3-6}\cmidrule(lr){7-8}
    Checkpoint & Estimate read & Judge \(\uparrow\) & J-swap \(\uparrow\) & CLAP \(\uparrow\) & C-swap \(\uparrow\) & WER \(\downarrow\) & Cosine \(\downarrow\)\\
    \midrule
    Generation & below \(\sigma_g{=}0.97\) (ours) & \(3.82\pm0.01\) & \(\mathbf{2.07\pm0.01}\) & \(\underline{0.15\pm0.00}\) & \(\underline{0.18\pm0.00}\) & \(\mathbf{0.07\pm0.03}\) & \(\mathbf{0.66\pm0.01}\)\\
    Generation & every step & \(3.79\pm0.05\) & \(1.91\pm0.14\) & \(\underline{0.15\pm0.00}\) & \(\underline{0.18\pm0.00}\) & \(\mathbf{0.07\pm0.02}\) & \(\mathbf{0.66\pm0.01}\)\\
    Generation & never & \(\mathbf{3.90\pm0.04}\) & \(1.93\pm0.06\) & \(\mathbf{0.18\pm0.00}\) & \(\mathbf{0.21\pm0.00}\) & \(0.16\pm0.02\) & \(\mathbf{0.66\pm0.00}\)\\
    Separation & every step & \(\underline{3.83\pm0.01}\) & \(\underline{1.95\pm0.13}\) & \(0.14\pm0.00\) & \(\underline{0.18\pm0.00}\) & \(\underline{0.10\pm0.03}\) & \(\underline{0.68\pm0.01}\)\\
    \bottomrule
  \end{tabular}%
  }
\end{table}

\FloatBarrier
\section{Separation results}
\label{sec:septables}

Table~\ref{tab:veo} reports the full Veo benchmark grid with
paired-test stars, and Tables~\ref{tab:ltx} and~\ref{tab:sam172} the
LTX-2.3 benchmark and SAM-Audio-bench. Both checkpoints lead every
cascade baseline on every column of all three benchmarks, except CLAP
identity on Veo, where AudioSep, whose query encoder is the scorer's
CLAP checkpoint, ties the separation checkpoint.

\begin{table*}[!htb]
  \centering
  \caption{Veo benchmark (sounds \(n{=}104\), speech \(n{=}45\)) with seed standard deviations and the paired-test significance of \method{} (\ckptsep{}) against each baseline. The bottom block applies SAM~Audio's best-of-eight protocol to both methods (Section~\ref{sec:candidates}); its stars compare with \method{} at eight candidates, and rankings are computed over the single-draw rows.}
  \label{tab:veo}\label{tab:k8veo}
  \footnotesize
  \setlength{\tabcolsep}{3.5pt}
  \fitwidth{%
  \begin{tabular}{l rr rr rr}
    \toprule
    & \multicolumn{4}{c}{Sounds (\(n{=}104\))}
    & \multicolumn{2}{c}{Speech (\(n{=}45\))} \\
    \cmidrule(lr){2-5}\cmidrule(lr){6-7}
        Method & Judge \(\uparrow\) & J-swap \(\uparrow\) & CLAP \(\uparrow\) & C-swap \(\uparrow\) & WER \(\downarrow\) & Cosine \(\downarrow\)\\
    \midrule
        SAM large & \(3.14\pm 0.22^{***}\) & \(0.50\pm 0.15^{***}\) & \(0.07\pm 0.02^{***}\) & \(0.04\pm 0.01^{***}\) & \(\underline{0.74\pm 0.06}^{***}\) & \(0.95\pm 0.02^{***}\)\\
        SAM large-tv & \(3.06\pm 0.06^{***}\) & \(0.35\pm 0.02^{***}\) & \(0.07\pm 0.01^{***}\) & \(0.04\pm 0.00^{***}\) & \(0.81\pm 0.02^{***}\) & \(0.96\pm 0.01^{***}\)\\
        FlowSep & \(3.21\pm 0.13^{***}\) & \(1.33\pm 0.11^{***}\) & \(0.15\pm 0.01^{**}\) & \(0.10\pm 0.01^{***}\) & \(0.98\pm 0.02^{***}\) & \(0.94\pm 0.00^{***}\)\\
        AudioSep & \(3.46^{***}\) & \(1.60^{***}\) & \(\underline{0.19}^{\mathrm{ns}}\) & \(\underline{0.16}^{***}\) & \(0.93\pm 0.00^{***}\) & \(\underline{0.89}^{***}\)\\
        SA\(\times\)SAM3 & \(3.07\pm 0.02^{***}\) & \(0.27\pm 0.07^{***}\) & \(0.08\pm 0.01^{***}\) & \(0.03\pm 0.01^{***}\) & \(0.87\pm 0.04^{***}\) & \(0.94\pm 0.01^{***}\)\\
    \midrule
        \system{} (\ckptsep{}) & \(\underline{4.36\pm 0.01}\) & \(\mathbf{2.61\pm 0.01}\) & \(\underline{0.19\pm 0.01}\) & \(\mathbf{0.23\pm 0.00}\) & \(\mathbf{0.05\pm 0.01}\) & \(\mathbf{0.61\pm 0.00}\)\\
    \system{} (\ckptgen{}) & \(\mathbf{4.39\pm 0.01}\) & \(\underline{2.59\pm 0.03}\) & \(\mathbf{0.20\pm 0.00}\) & \(\mathbf{0.23\pm 0.00}\) & \(\mathbf{0.05\pm 0.01}\) & \(\mathbf{0.61\pm 0.00}\)\\
    \midrule
    \multicolumn{7}{l}{\emph{Best of eight candidates, SAM~Audio ranker}}\\
        SAM large & \(4.10^{**}\) & \(1.81^{***}\) & \(0.16^{***}\) & \(0.15^{***}\) & \(0.56^{***}\) & \(0.77^{***}\)\\
        SAM large-tv & \(4.15^{**}\) & \(1.82^{***}\) & \(0.17^{**}\) & \(0.15^{***}\) & \(0.55^{***}\) & \(0.77^{***}\)\\
        \system{} (\ckptsep{}) & \(4.49\) & \(2.70\) & \(0.20\) & \(0.24\) & \(0.05\) & \(0.61\)\\
    \bottomrule
  \end{tabular}%
  }
\end{table*}

\paragraph{Candidate count.}\label{sec:candidates} Under SAM~Audio's eight-candidate
protocol (Section~\ref{sec:separation} of the main paper), the ranker selects on Judge
and CLAP, two of the measures reported here. We regard single draws as the
fairer comparison: reranking is a selection step that is independent of the
separator and could be added to any method, and because it selects on Judge
and CLAP it raises those columns by construction. We nonetheless report both
settings, and \method{} matches or surpasses the baselines in each. Apart from the bottom block of Table~\ref{tab:veo}, all tables report each method's direct output; that block
evaluates both methods under SAM~Audio's protocol on the
Veo benchmark, with the same captions and the
same reranker applied to eight independent draws of each. Reranking
changes little for \method{}, whose eight draws have a Judge
standard deviation of 0.02, and much for both SAM~Audio checkpoints,
whose candidates range from clean to silent. After reranking
\method{} keeps a significant lead over both checkpoints on every
measure, most clearly on those the reranker does not select on,
own-line word error and speaker cosine.

\begin{table*}[!htb]
  \centering
  \caption{LTX-2.3 benchmark (sounds \(n{=}37\), speech \(n{=}27\)). SA\(\times\)SAM3 stems obtained from empty masks are silent and are included in all means.}
  \label{tab:ltx}
  \footnotesize
  \setlength{\tabcolsep}{3.5pt}
  \fitwidth{%
  \begin{tabular}{l rr rr rr}
    \toprule
    & \multicolumn{4}{c}{Sounds (\(n{=}37\))}
    & \multicolumn{2}{c}{Speech (\(n{=}27\))} \\
    \cmidrule(lr){2-5}\cmidrule(lr){6-7}
        Method & Judge \(\uparrow\) & J-swap \(\uparrow\) & CLAP \(\uparrow\) & C-swap \(\uparrow\) & WER \(\downarrow\) & Cosine \(\downarrow\)\\
    \midrule
        SAM large & \(3.40\pm 0.10\) & \(0.85\pm 0.05\) & \(0.11\pm 0.02\) & \(0.07\pm 0.00\) & \(0.78\pm 0.01\) & \(0.96\pm 0.01\)\\
        SAM large-tv & \(3.06\pm 0.19\) & \(0.39\pm 0.25\) & \(0.11\pm 0.01\) & \(0.03\pm 0.01\) & \(0.88\pm 0.01\) & \(0.95\pm 0.01\)\\
        FlowSep & \(3.15\pm 0.11\) & \(1.00\pm 0.10\) & \(0.15\pm 0.01\) & \(0.07\pm 0.01\) & \(1.08\pm 0.06\) & \(0.94\pm 0.04\)\\
        AudioSep & 2.92 & 1.01 & \underline{0.16} & 0.11 & 0.98 & \underline{0.90}\\
        SA\(\times\)SAM3 & \(2.84\pm 0.18\) & \(0.52\pm 0.19\) & \(0.10\pm 0.01\) & \(0.03\pm 0.01\) & \(0.90\pm 0.05\) & \(0.96\pm 0.02\)\\
    \midrule
        \system{} (\ckptsep{}) & \(\mathbf{4.08\pm 0.03}\) & \(\mathbf{2.04\pm 0.04}\) & \(\mathbf{0.19\pm 0.00}\) & \(\mathbf{0.18\pm 0.01}\) & \(\mathbf{0.09\pm 0.01}\) & \(\mathbf{0.66\pm 0.01}\)\\
        \system{} (\ckptgen{}) & \(\underline{4.01\pm 0.03}\) & \(\underline{1.82\pm 0.03}\) & \(\mathbf{0.19\pm 0.00}\) & \(\underline{0.16\pm 0.00}\) & \(\underline{0.13\pm 0.01}\) & \(\mathbf{0.66\pm 0.01}\)\\
    \bottomrule
  \end{tabular}%
  }
\end{table*}

\begin{table}[!htb]
  \centering
  \caption{SAM-Audio-bench (172 clips). Each baseline is queried both with our captions (\emph{ours}) and with the native benchmark text (\emph{native}); rankings are computed over the \emph{ours} rows.}
  \label{tab:sam172}
  \small
  \setlength{\tabcolsep}{4pt}
  \begin{tabular}{ll rrrr}
    \toprule
        Method & Captions & Judge \(\uparrow\) & J-swap \(\uparrow\) & CLAP \(\uparrow\) & C-swap \(\uparrow\)\\
    \midrule
        SAM large-tv & ours & \(\underline{3.12\pm 0.08}\) & \(0.78\pm 0.14\) & \(0.13\pm 0.01\) & \(0.03\pm 0.01\)\\
         & native & \(3.32\pm 0.11\) & \(0.98\pm 0.12\) & \(0.14\pm 0.01\) & \(0.04\pm 0.01\)\\
    \midrule
        AudioSep & ours & \(2.81\pm 0.00\) & \(0.82\pm 0.00\) & \(\underline{0.20\pm 0.00}\) & \(0.07\pm 0.00\)\\
         & native & \(2.38\pm 0.00\) & \(0.51\pm 0.00\) & \(0.20\pm 0.00\) & \(0.03\pm 0.00\)\\
    \midrule
        SAM large & ours & \(2.75\pm 0.15\) & \(0.63\pm 0.06\) & \(0.10\pm 0.01\) & \(0.01\pm 0.01\)\\
         & native & \(2.70\pm 0.13\) & \(0.58\pm 0.09\) & \(0.10\pm 0.01\) & \(0.01\pm 0.01\)\\
    \midrule
        FlowSep & ours & \(2.64\pm 0.06\) & \(0.51\pm 0.08\) & \(0.16\pm 0.01\) & \(0.03\pm 0.01\)\\
         & native & \(2.29\pm 0.10\) & \(0.33\pm 0.07\) & \(0.17\pm 0.00\) & \(0.00\pm 0.01\)\\
    \midrule
        \system{} (\ckptsep{}) & ours & \(\mathbf{3.70\pm 0.06}\) & \(\underline{1.42\pm 0.06}\) & \(\mathbf{0.21\pm 0.00}\) & \(\underline{0.11\pm 0.01}\)\\
        \system{} (\ckptgen{}) & ours & \(\mathbf{3.70\pm 0.06}\) & \(\mathbf{1.52\pm 0.06}\) & \(\mathbf{0.21\pm 0.00}\) & \(\mathbf{0.13\pm 0.01}\)\\
    \bottomrule
  \end{tabular}
\end{table}

\paragraph{Speech separators.} Table~\ref{tab:speechsep} compares
\system{} with two pretrained speech separators, MossFormer2
\citep{Zhao2023MossFormer2CT} and SepFormer \citep{Subakan2020AttentionIA}
trained on Libri2Mix \citep{cosentino2020librimixopensourcedatasetgeneralizable}. Each separates the
audio-mix into two outputs, which are assigned to the two captions by
matching their transcripts to the quoted lines. \system{} has lower
own-line word error on all three benchmarks; on Veo, SepFormer's
speaker cosine is 0.59 against 0.61, not significant.

\begin{table}[!htb]
  \centering
  \caption{Comparison with dedicated speech separators on the three speech lanes. Separator outputs are assigned to captions by matching their transcripts to the quoted lines.}
  \label{tab:speechsep}
  \footnotesize
  \setlength{\tabcolsep}{3.5pt}
  \fitwidth{%
  \begin{tabular}{l rr rr rr}
    \toprule
    & \multicolumn{2}{c}{Generation} & \multicolumn{2}{c}{Veo} & \multicolumn{2}{c}{LTX-2.3}\\
    \cmidrule(lr){2-3}\cmidrule(lr){4-5}\cmidrule(lr){6-7}
    Method & WER \(\downarrow\) & Cosine \(\downarrow\) & WER \(\downarrow\) & Cosine \(\downarrow\) & WER \(\downarrow\) & Cosine \(\downarrow\)\\
    \midrule
    MossFormer2 & \(0.90^{***}\) & \(0.77^{**}\) & \(0.68^{***}\) & \(0.65^{\mathrm{ns}}\) & \(0.85^{***}\) & \(0.74^{*}\)\\
    SepFormer (Libri2Mix) & \(0.58^{***}\) & \(0.74^{**}\) & \(0.53^{***}\) & \(\mathbf{0.59}^{\mathrm{ns}}\) & \(0.39^{**}\) & \(0.73^{*}\)\\
    \midrule
    \system{} & \(\mathbf{0.07\pm 0.03}\) & \(\mathbf{0.66\pm 0.01}\) & \(\mathbf{0.05\pm 0.01}\) & \(0.61\pm 0.01\) & \(\mathbf{0.09\pm 0.01}\) & \(\mathbf{0.66\pm 0.01}\)\\
    \bottomrule
  \end{tabular}%
  }
\end{table}

\paragraph{Audio-mix conditioning.} Table~\ref{tab:mixablation}
replaces the audio-mix channel of the separation checkpoint on the
LTX-2.3 clip set: a silent audio-mix collapses every stem, whereas
a cross-clip donor leaves the stems live and separating the donor
scene, transcribing the donor's lines rather than the caption's
script, the inverse of the observed-audio-mix run, so the stems
decompose the audio-mix they observe.

\begin{table}[!htb]
  \centering
  \caption{Audio-mix ablation on the LTX-2.3 clip set (\ckptsep{}, 64 clips), in which the observed audio-mix is replaced by silence or by the audio-mix of another clip.}
  \label{tab:mixablation}
  \small
  \setlength{\tabcolsep}{5pt}
  \begin{tabular}{lccc}
    \toprule
    Audio-mix clamped to & Judge \(\uparrow\)
      & \multicolumn{2}{c}{Word error}\\
    \cmidrule(lr){3-4}
    & & Own-line \(\downarrow\) & Donor-line\\
    \midrule
    Observed scene & \textbf{3.80} & \textbf{0.09} & 0.99\\
    Cross-clip donor & \underline{2.97} & 1.16 & 0.41\\
    Digital silence & 1.10 & \underline{1.00} & --\\
    \bottomrule
  \end{tabular}
\end{table}

\FloatBarrier
\section{Waveform-level evaluation}

\paragraph{Reference-stem evaluation.}
\label{sec:refstem}
Every training family is additive, so held-out recordings from
them carry true stems. We hold out 129 two-source scenes at native
levels, 60 URMP duets from pieces absent from training (26 with two
of the same instrument), 57 CelebV-HQ dialogues from the
identity-disjoint split, and 12 validation composites, 258 stems in
all; URMP and CelebV-HQ are outside the training lists of AudioSep
and SAM~Audio. Each output stem is compared with its reference
after level matching and a \(\pm50\)\,ms lag search, by 20\,ms
envelope correlation and 64-band log-spectral distance (LSD), the
two measures that survive the audio VAE decoder (see the SI-SDR
paragraph and Table~\ref{tab:sisdr}); controls are the audio-mix in both
stems, the true stems and each cascade baseline's stems after one
decoder pass, and a cross-clip null. \method{} leads on both
measures (Table~\ref{tab:refstem}; LSD \(6.39\pm0.05\) over three
seeds), including over SAM~Audio at eight ranker-selected draws. On the 57
dialogues our stems transcribe at the own-line word error of the
true stems, 0.05, where the strongest cascade baseline reads 0.29
and the audio-mix 1.03. Same-instrument duets carry identical
captions on both stems; there \method{} reads 11.7\,dB LSD against
6.3\,dB on distinct-instrument duets, with permutation-invariant
LSD (11.4\,dB) close to caption-fixed LSD, so the two stems are a
split that the caption cannot direct, and no permutation fixes it.

\begin{table}[!htb]
  \centering
  \caption{Reference-stem evaluation on 129 held-out scenes with ground-truth stems (258 stems, seed 0), measured by envelope correlation and log-spectral distance. Best-of-eight rows and controls are not ranked.}
  \small
  \setlength{\tabcolsep}{4pt}
  \begin{tabular}{lcc}
    \toprule
    Method & Envelope \(r\) \(\uparrow\) & LSD \(\downarrow\)\\
    \midrule
    \method{} (\ckptsep{}) & \textbf{0.82} & \textbf{6.3}\\
    AudioSep & \underline{0.76} & \underline{11.0}\\
    FlowSep & 0.65 & 13.1\\
    SAM large-tv & 0.47 & 17.6\\
    SAM large & 0.39 & 18.9\\
    SAM large-tv, best of eight & 0.78 & 10.3\\
    SAM large, best of eight & 0.78 & 10.3\\
    \midrule
    Audio-mix in both stems & 0.69 & 12.4\\
    True stems, VAE round trip & 0.96 & 3.1\\
    AudioSep, VAE round trip & 0.73 & 11.6\\
    SAM large-tv, best of eight, round trip & 0.76 & 10.6\\
    Cross-clip null & 0.04 & 23.8\\
    \bottomrule
  \end{tabular}
  \label{tab:refstem}
\end{table}

\paragraph{SI-SDR and the audio VAE.}
\label{sec:sisdr}

SI-SDR compares a stem with a reference sample by sample. It suits
mask-based separators such as AudioSep, which filter the audio-mix.
Generative separators synthesize a new waveform. Our audio VAE
decoder, for example, regenerates the fine time structure of each
stem from its latent. Passing each method's stems once
through the decoder and scoring every input against its round trip
(Table~\ref{tab:sisdr}) makes SI-SDR negative for every method
while envelope correlation stays high and LSD low: the decoder
reproduces energy and spectrum while rewriting the waveform.
Restoring the original short-time phase alone makes SI-SDR positive
again, whereas restoring the original magnitudes alone barely moves
it, so the loss lies in the resynthesized phase and we score
separation with phase-invariant measures. The bottom block of
Table~\ref{tab:sisdr} scores each method's summed stems against its
audio-mix by the same measures; a low SI-SDR there is expected for
decoder-resynthesized stems, and envelope correlation and LSD are
the operative columns.

\begin{table}[!htb]
  \centering
  \caption{Audio VAE decoder round trip (194 clips, seed 0). Top: the stems of each method evaluated against their own round trip. Bottom: the summed stems evaluated against the audio-mix.}
  \small
  \setlength{\tabcolsep}{4pt}
  \begin{tabular}{lccccc}
    \toprule
    & \multicolumn{3}{c}{Stem vs.\ round trip}
      & \multicolumn{2}{c}{SI-SDR (dB), restoring original}\\
    \cmidrule(lr){2-4}\cmidrule(lr){5-6}
    Method & SI-SDR (dB) & envelope \(r\) & LSD (dB) & phase & magnitudes\\
    \midrule
    FlowSep & \(-25.2\) & 0.99 & 2.8 & \(+15.3\) & \(-25.0\)\\
    AudioSep & \(-13.5\) & 0.99 & 2.6 & \(+15.7\) & \(-13.2\)\\
    SAM large-tv & \(-26.1\) & 0.78 & 2.0 & \(+6.5\) & \(-24.5\)\\
    SAM large & \(-25.1\) & 0.81 & 2.1 & \(+7.5\) & \(-23.8\)\\
    \method{} (2nd round trip) & \(-10.4\) & 0.96 & 2.2 & \(+15.8\) & \(-9.7\)\\
    \midrule
    \multicolumn{6}{l}{\emph{Stem sum vs.\ its audio-mix} (\(n{=}194\) clips per method)}\\
    \midrule
    FlowSep & \(-28.5\) & 0.82 & 9.1 & -- & --\\
    AudioSep & \(+12.7\) & 0.84 & 6.9 & -- & --\\
    SAM large-tv & \(-16.0\) & 0.50 & 25.2 & -- & --\\
    SAM large & \(-14.9\) & 0.47 & 26.5 & -- & --\\
    \method{} & \(-17.0\) & 0.92 & 4.8 & -- & --\\
    \bottomrule
  \end{tabular}
  \label{tab:sisdr}
\end{table}

\FloatBarrier
\section{Placement readout}
\label{sec:locapp}

\begin{figure*}[htbp]
  \centering
  \includegraphics[width=0.78\linewidth]{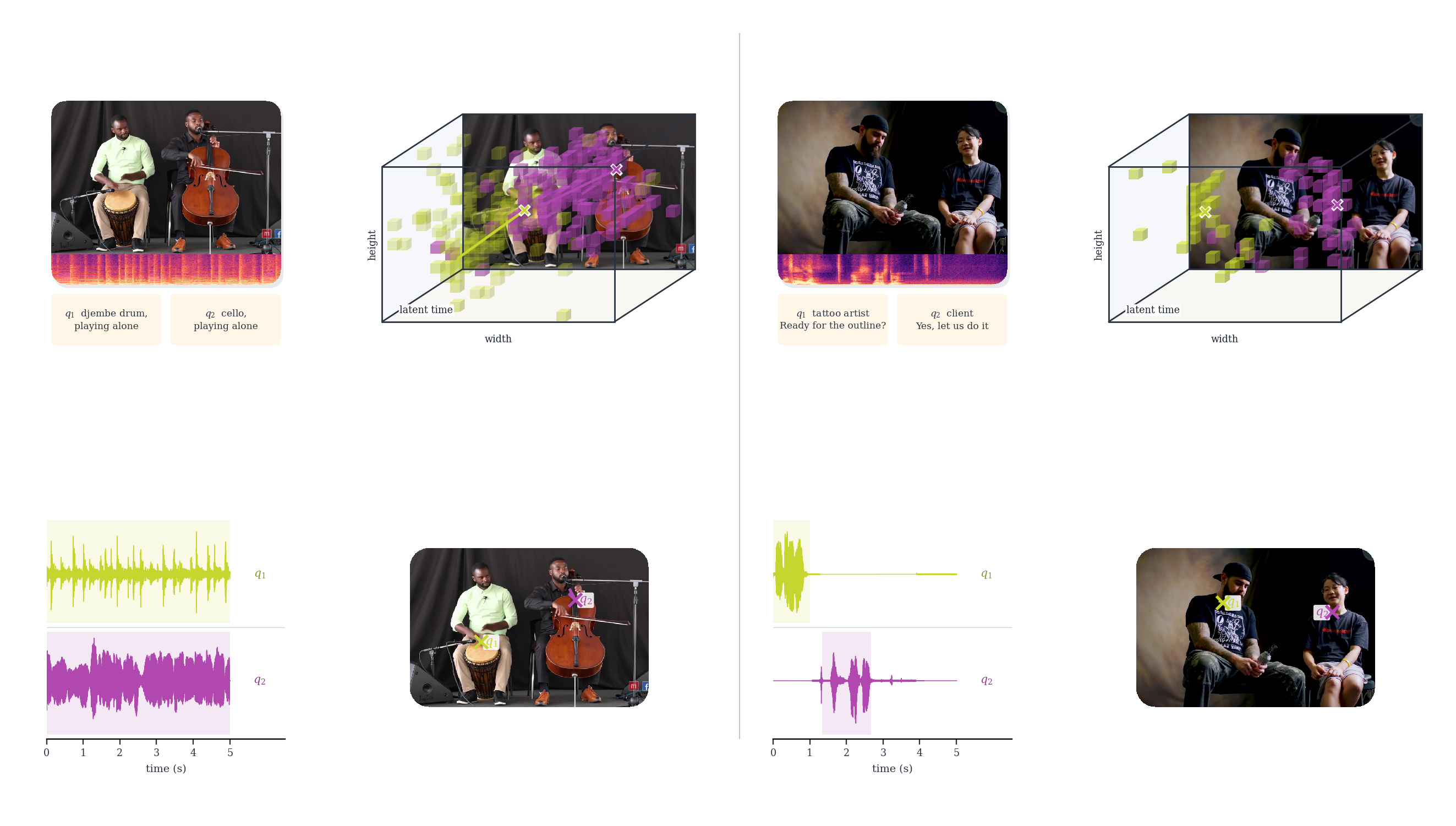}
  \caption{\textbf{Same-pass placement readout on two scenes.}
  Left: instrument couple (djembe / cello).
  Right: turn-taking dialogue (tattoo artist / client).
  Each column shows the observed video and mix, the two captions, the
  \(H{\times}W{\times}T\) attention volume (16 latent frames), the two source
  waveforms, and the pooled geometric-median + Kalman tracks. Voxels and tracks
  are drawn only on latent bins where that source waveform is active; quiet bins
  are omitted. Colors match the two captions.}
  \label{fig:avvolume}
\end{figure*}

We track each source in the image from the attention of its stem,
with no extra training or detector. Each stem has two attention maps:
one between the video and the words of its caption, and one from the
video to the stem's audio. We average both over the 30 separation
steps and blocks 26 to 33 and 36, and keep the map that gives the
more stable track. In each frame we take the weighted geometric
median of its strongest 2\% and smooth the track with a Kalman
smoother. Earlier work finds sounding objects with audio-visual
attention \citep{senocak2018learning} or by separating a sound for
each pixel \citep{zhao2018soundofpixels}. Figure~\ref{fig:avvolume}
shows the tracks.

\FloatBarrier

\end{document}